\documentclass{JHEP}
\usepackage{epsfig}
\usepackage{amssymb,latexsym,amsfonts}

\newcommand{\be}{\begin{equation}}
\newcommand{\ee}{\end{equation}}
\newcommand{\bd}{\begin{displaymath}}
\newcommand{\ed}{\end{displaymath}}
\newcommand{\ba}{\begin{array}}
\newcommand{\ea}{\end{array}}
\newcommand{\bt}{\begin{tabular}}
\newcommand{\et}{\end{tabular}}

\newcommand{\bea}{\begin{eqnarray}}
\newcommand{\eea}{\end{eqnarray}}
\newcommand{\bean}{\begin{eqnarray*}}
\newcommand{\eean}{\end{eqnarray*}}

\newcommand{\hlf}{\frac{1}{2}}

\newcommand{\dif}{\mathrm{d}}

\newcommand{\inp}[2]{\langle #1, #2 \rangle}

\newcommand{\Z}{\mathbb{Z}}

\newcommand{\R}{\mathbb{R}}
\newcommand{\C}{\mathbb{C}}

\newcommand{\mod}{\textrm{mod}}

\newcommand{\df}[1]{{} * \! #1}

\preprint{{\tt hep-th/0404174}}
\title{Time-like T-duality algebra}

\author{Arjan Keurentjes
\\ {\it Theoretische Natuurkunde, Vrije Universiteit Brussel, and\\
 the International Solvay Institutes,\\Pleinlaan 2, B-1050 Brussels, Belgium}
\\ \email{arjan@tena4.vub.ac.be} }

\abstract{When compactifying $M$- or type $II$ string-theories on tori
 of indefinite space-time signature, their low energy theories involve sigma
 models on $E_{n(n)}/H_n$, where $H_n$ is a not necessarily compact
 subgroup of $E_{n(n)}$ whose complexification is identical to the
 complexification of the maximal compact subgroup of $E_{n(n)}$. We
 discuss how to compute the group $H_n$. For finite dimensional
 $E_{n(n)}$, a formula derived from the theory of real forms of $E_n$
 algebra's gives the possible groups immediately. A few groups
 that have not appeared in the literature are found. For $n=9,10,11$
 we compute and describe the relevant real forms of $E_n$
 and $H_n$. A given $H_n$ can correspond
 to multiple signatures for the compact torus. We compute the groups
 $H_n$ for all compactifications of $M$-, $M^*$-, and $M'$-theories,
 and type $II$- $II^*$- and $II'$- theories on tori of arbitrary
 signature, and collect them in tables that outline the dualities
 between them. In an appendix we list cosets $G/H$, with $G$ split and
 $H$ a subgroup of $G$, that are relevant to timelike toroidal
 compactifications and oxidation of theories with enhanced
 symmetries.} \keywords{Space-time symmetries; M-theory; String-duality} 

\begin{document}
\section{Introduction}

The advent and understanding of duality symmetries has radically
changed our view of high-energy physics. Such symmetries transform
excitations of elementary fields into solitons, can change the
topology of space-time, and allows to interpret seemingly very
different theories, as different expansions from a single underlying
theory \cite{Hull:1994ys, Witten:1995ex}.  

Under duality symmetries not much is sacred, and it was found that
there are even duality symmetries that can change the signature of
space-time \cite{Hull:1998vg, Hull:1998ym, Hull:1998fh}. To establish
such a duality, one compactifies the theory on a time-like
circle. There are however a number of subtleties with time-like
compactifications. 

It has been known for a long time that toroidal
compactification of 11 dimensional supergravity \cite{Cremmer:1978km}
results in effective theories with the scalars organized in sigma models
on symmetric spaces of the form $E_{n(n)}/H_n$ \cite{Cremmer:1979up,
  Julia:1980gr, Julia:1982gx} where $E_{n(n)}$ is the split (maximal
non-compact) real form of an exceptional Lie-algebra, and $H_n$ is its
maximal compact subgroup. The number $n$ equals $11-d$ for a
compactification to $d$ dimensions. For $d \geq 3$ the global symmetry
$E_{n(n)}$ has been checked in detail in \cite{Cremmer:1997ct}, which for
$d=4$ was already established in \cite{Cremmer:1979up} (see
\cite{Obers:1998fb} for a review and more references). For $0 <d < 3$
the symmetries are supported by considerable evidence
\cite{Julia:1982gx, Nicolai:kz, Nicolai:1988jb, Nicolai:1998gi,
  Damour:2002cu}.    

Applying the reduction program to include time-like directions,
entails a number of modifications. The coset symmetries found still
have scalars on cosets $E_{n(n)}/H_n$, but although the denominator
group $E_{n(n)}$ is still the split real form of the exceptional $E_n$
group, the groups $H_n$ are no longer compact \cite{Hull:1998br,
  Cremmer:1998em}. Instead the groups $H_n$ have the same
complexification as the maximal compact subgroup, but typically are
non-compact real forms. The papers \cite{Hull:1998br, Cremmer:1998em}
considered dimensional reduction of conventional supergravities (in a
space-time with 1 time-direction) with \emph{one} time-direction
included, yet the arguments of \cite{Hull:1998vg, Hull:1998ym,
  Hull:1998fh} imply the existence of less conventional supergravities
formulated on space-times with more than one time-like direction. 

In this paper, we study all these variant supergravities, reduced over
tori of arbitrary space-time signature. There are many possibilities
(many supergravity variants, many possibilities for the space-time
signature), and it is actually more convenient to study the problem
from the other end. The global symmetry groups $E_{n(n)}$ are constant
elements in the discussion, so we need to determine which groups $H_n$
can appear as denominator subgroups, and, how these groups relate to
the space-time signature of the compact torus, as well as the various
signs for the form-field terms in the Lagrangian that distinguish
various supergravities from one another.

This is possible using extensions of the techniques from
\cite{Keurentjes:2004bv}. In this paper the implications of
space-time signature in the context of the $E_{11}$-conjecture
\cite{West:2001as} were studied. The $E_{11}$-conjecture states that
there exists a hypothetical formulation of (the bosonic sector of)
\emph{11-dimensional} supergravity/$M$-theory, with a non-linearly realized
$E_{11}$ symmetry. The variables describing the theory are specified
by the coset $E_{11(11)}/H_{11}$, where $H_{11}$ is a subgroup of
$E_{11}$. This subgroup is necessarily non-compact, as it is proposed
that $H_{11}$ contains the Lorentz-group $SO(1,10)$ for 11
dimensions. The paper \cite{Keurentjes:2004bv} demonstrates that there
are also other real forms of $SO(11,\C)$ contained in $H_{11}$. The
possibility to select these as Lorentz group implies that a formalism based on
a non-linearly realized $E_{11}$ symmetry (if true) describes not only
conventional 11 dimensional supergravity, but also
theories in other space-time signatures. In this way it makes contact
with the results of \cite{Hull:1998vg, Hull:1998ym}, and indeed one
can show that the $E_{11}$-conjecture has the potential to describe
all the theories in these papers.

The techniques that were applied in \cite{Keurentjes:2004bv} to
$E_{11(11)}$ and $H_{11}$ apply equally well to $E_{n(n)}$ and $H_n$
for $n \neq 11$ (and in fact to any Lie group $G$). Two crucial
elements in the discussion of \cite{Keurentjes:2004bv} are
$\Z_2$-valued functions on the root lattice and Weyl-reflections. As we will
explain in this paper, the classification of $\Z_2$-valued functions
amounts to nothing but the classification of inner involutions on the
$E_n$ algebra. For the finite dimensional $E_n$ this has been
established long ago, in the context of the classification of real
forms of the algebra. Working out this connection, we obtain a simple
formula which immediately gives all the possible denominator
subgroups. Among these there are a few that have not appeared in
\cite{Hull:1998br, Cremmer:1998em, Hull:1998vg, Hull:1998ym}. For
infinite-dimensional $E_n$ and $H_n$ our techniques fix the
real forms that are possible in principle, and those that actually do
appear in compactifications of maximal supergravities (just as in the
$E_{11}$ case, a number of real forms is ruled out by the requirement
that they have to connect to conventional 11 dimensional supergravity
\cite{Keurentjes:2004bv}). 

Weyl reflections are instrumental. Establishing the global and local
symmetries in supergravity, it is conventional to choose a particular
realization of the symmetry, with a one-to-one correspondence between
the basis of the Cartan subalgebra of $E_{n(n)}$ and the dilatonic
scalars in the theory (see e.g. \cite{Cremmer:1997ct, Cremmer:1999du,
  Keurentjes:2002xc, Keurentjes:2002rc}). The roots of the algebra,
and their associated operators then correspond to axions and their
duals. In this form the algebra is essentially fixed, and there are no
continuous $E_n$ transformations possible anymore. There is however a
discrete set of $E_n$ transformations that rotate the root lattice
into itself; these are the Weyl-reflections. As we will recall in
subsection \ref{Weyl} these Weyl-reflections correspond to nothing but
a particular subset of the duality transformations in the theory; this
implies that the action of the Weyl-group is closely linked to the group of
T-dualities, including the time-like ones.

The set-up of this paper is as follows. In section \ref{group} we set
the stage for the mathematical formalism and introduce our
conventions. We then explain how our problem is linked to the theory
of real forms of algebra's, culminating in equation (\ref{sig}) which
is one of our core results. Section \ref{comp} explains how to perform explicit
computations, identify dualities, extract space-time signatures and
other signs. Section \ref{Msmall} applies the formalism to
compactifications of $M$-theory to $d \geq 3$ dimensions, where we can
use the link to real forms of $E_n$ algebra's to shortcut a number of
computations. Section \ref{Mlarge} deals with the same problem, but
now for $d < 3$. As the mathematical groups appearing here are less
familiar, we devote some more space to discussion of their
properties. Section \ref{IIA} deals with the (relatively trivial) link
to $IIA$-string and supergravity theories. Section \ref{IIB} studies
an alternative embedding of the Lorentz-algebra of the torus,
resulting in $IIB$-theories and their variants. The sections
\ref{Msmall}, \ref{Mlarge}, \ref{IIB} contain many tables that should
be helpful to the reader who is interested in the dualities and the
cosets, but not necessarily in the full machinery behind
them. Finally, in section \ref{conc}, we summarize our results. We
have added an appendix with a table containing the groups related by
our equation (\ref{sig}) for cosets with a split Lie-group in the
numerator; these are relevant for time-like reduction and oxidation of
theories described in \cite{Cremmer:1999du, Keurentjes:2002xc}.
  
\section{Group theory} \label{group}

\subsection{Definition and properties of the $E$-algebra's}

In this section we recall some facts about the general theory of
Kac-Moody algebra's \cite{Kac:gs}, and the $E_n$ algebra's in
particular. Our conventions are chosen such that for $n=11$ we recover the
conventions of the paper \cite{Keurentjes:2004bv}, apart from the fact
that here we order the nodes along the horizontal line in the Dynkin
diagram in the opposite direction.

We start by drawing the Dynkin diagrams for $E_n$.

\FIGURE{
\includegraphics[width=13cm]{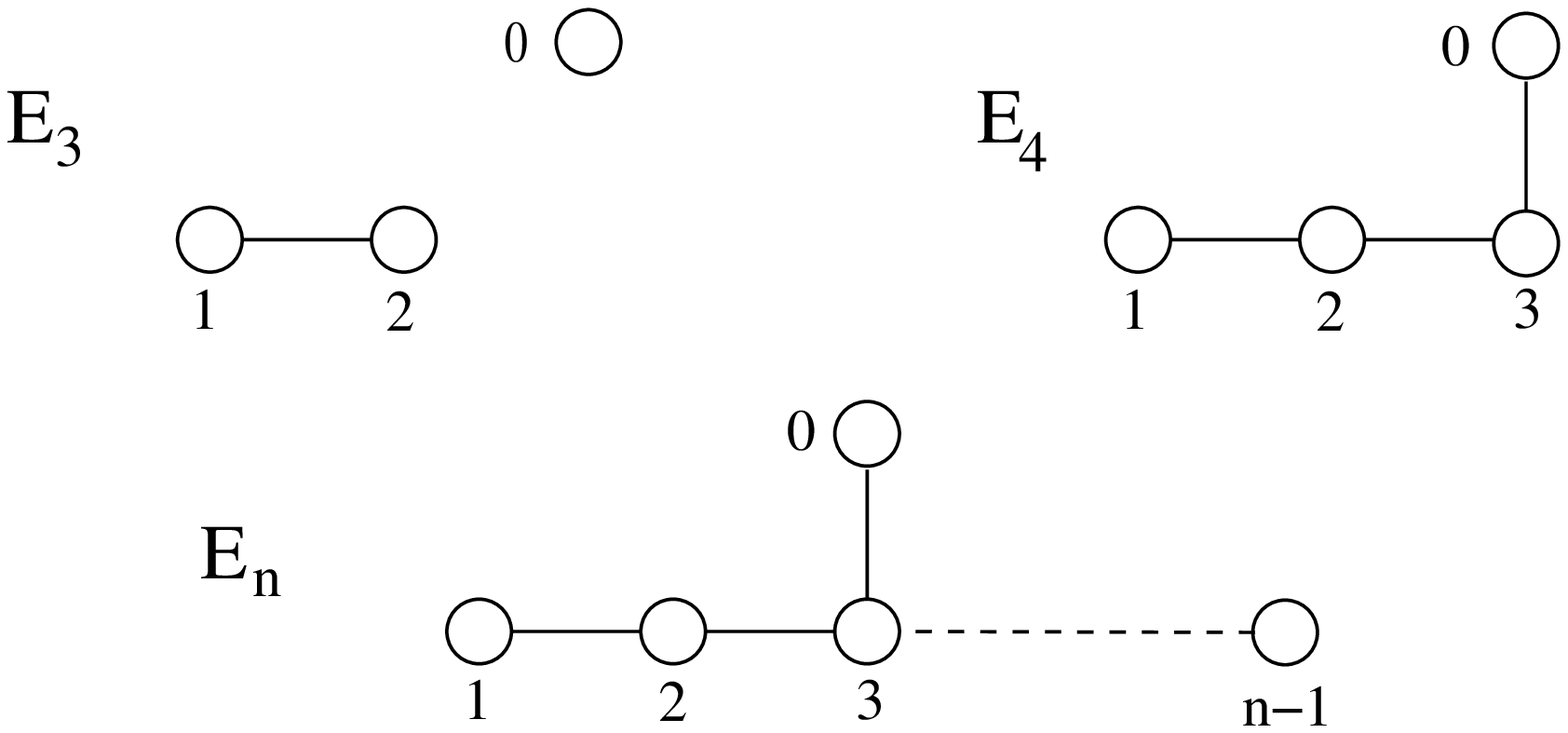}
\caption{The Dynkin diagrams of $E_{n}$ algebra's.}\label{e11fig}
}

For $n <6$ there exist alternative names for these algebra's, as $E_3
\cong A_1 \oplus A_2$, $E_4 \cong A_4$, $E_5 \cong D_5$.

From the Dynkin diagram the Cartan matrix $A_n=(a_{ij})$, with $i,j$ in the
index set $I \equiv \{0, 1, \ldots, n-1 \}$, may be reconstructed by setting
\be
a_{ij} \equiv \left\{
\ba{rl}
 2 & \textrm{if }i=j; \\
-1 & \textrm{if }i,j\textrm{ connected by a line;}\\
 0 & \textrm{otherwise.}\\
\ea
\right.
\ee
The Cartan matrix is symmetric, and $\det(A_n)= 9-n$. For $n \neq 9$,
we choose a real vector space ${\cal H}$ of dimension $n$ and linearly
independent sets $\Pi = \{ \alpha_0, \ldots,\alpha_{n} \} \subset {\cal H}^*$
(with $\cal H^{*}$ the space dual to ${\cal H}$) and $\Pi^{\vee} = \{
\alpha_0^{\vee}, \ldots,\alpha_{n}^{\vee} \} \subset {\cal H}$,
obeying $a_{ij} = \alpha_j(\alpha_i^{\vee}) \equiv
\inp{\alpha_j}{\alpha_i^{\vee}}$. The elements of the set $\Pi$ are
called the simple roots. Linear combinations of the elements of $\Pi$
with integer coefficients span a lattice, called the root lattice of
$E_n$, which we will denote by $P_n$. Similarly, the $\alpha_i^{\vee}$
also span a lattice $P_n^{\vee}$ known as the coroot lattice.

For the affine Lie-algebra $E_9$ one has $\det(A_9)=0$, and the construction
of the full algebra requires one to introduce a vector space ${\cal
  H}$ of dimension 10 (instead of 9). The Cartan sub-algebra has one
extra generator, the central charge $K$. This special feature is
not playing any role in our discussion, as eventually we will be interested
in the subalgebra $H_9$ of $E_9$, of which $K$ is not a generator. We therefore
proceed, and refer to \cite{Kac:gs} for details on affine algebra's.

From the Cartan matrix the algebra $E_{n}$ can be constructed. The
generators of the algebra consist of 11 basis elements $h_i$ for the
Cartan sub algebra ${\cal H}$ together with 22 generators
$e_{\alpha_i}$ and $e_{-\alpha_i}$ ($i \in I$), and of algebra
elements obtained by taking multiple commutators of these. These
commutators are restricted by the algebraic relations (with
$h,h'\in{\cal H}$): 
\be
[h,h']=0 \quad [h, e_{\alpha_j}] = \inp{\alpha_{j}}{h} e_{\alpha_j}
\quad [h,e_{-\alpha_j}] = -\inp{\alpha_j}{h} e_{-\alpha_j} \quad
       [e_{\alpha_i}, e_{-\alpha_j}] =\delta_{ij} \alpha_i^{\vee}; 
\ee
and the Serre relations
\be
\textrm{ad}(e_{\alpha_i})^{1-a_{ij}} e_{\alpha_j} =0, \qquad
\textrm{ad}(e_{-\alpha_i})^{1-a_{ij}}e_{-\alpha_j} =0.
\ee
Although in the theories we will be considering the relevant real form of
$E_{n}$, is the so-called split real form, other real forms play an
auxiliary role in our computations. We will discuss real forms in
subsection \ref{realf}.

There is a natural bijection $\alpha_i \rightarrow \alpha_i^{\vee}$,
in which the components of each $\alpha_i^{\vee}$ turn out to be
identical to those of $\alpha_i$ (This is because the $E_{n}$ are simply-laced
algebra's). To some extent, one may regard the $\alpha_i^{\vee}$ as
``column vectors'', and the $\alpha_i$ as ``row vectors''.  Under the
bijection, the coroot lattice, consisting of linear combinations with integer
coefficients of $\alpha_i^{\vee}$ may therefore be identified with the
root lattice.
 
Under the root space decomposition with respect to the Cartan
subalgebra, $E_n$ is decomposed into subspaces of eigenvectors with
respect to the adjoint action of the Cartan
sub-algebra. Schematically, $E_{n} = \oplus_{\alpha \in {\cal H}^*}
g_{\alpha}$, with 
\be
g_{\alpha} = \{ x \in E_{n}: [h,x] = \inp{\alpha}{h} x \ \forall h
\in {\cal H} \}
\ee
The set of roots of the algebra, $\Delta$, are defined by
\be
\Delta = \{ \alpha \in {\cal H}: g_{\alpha} \neq 0, \alpha \neq 0  \}
\ee
The roots of the algebra form a subset of the root lattice $\Delta
\subset P_{n}$. The set of positive roots $\Delta^+ \subset \Delta$
is the subset of roots whose expansion in the simple roots
involves non-negative integer coefficients only. We denote the
basis elements of $g_{\alpha}$ by $e^k_{\alpha}$, where $k$ is a
degeneracy index, taking values in $\{1, \ldots, \dim(g_{\alpha})
\}$. If $\dim(g_{\alpha}) =1$, we will drop the degeneracy index, and
write $e_{\alpha}$ for the generator. Note that previously we defined
the generators $e_{\pm\alpha_i}$ for $\alpha_i$ a simple root, and as
$\dim(g_{\alpha_i}) =1$ when $\alpha_i$ is a simple 
root, this is consistent with the conventions we have laid out
here. With the aid of the Jacobi identity, it is easily proven that 
$[e^i_{\alpha}, e^j_{\beta}] \in g_{\alpha+\beta}$, if this commutator
is different from zero.

Another aspect that will enter prominently into the discussion are the
hermiticity properties of the operators $e^k_{\alpha}$ and $h$. In our
conventions these will be chosen as
\be \label{herm}
h^{\dag} = h \ \forall \ h \in {\cal H}  \qquad
\left(e^k_{\alpha}\right)^{\dag} = e^k_{-\alpha}.
\ee

For the specific $E_n$ root systems we recall that the set
$\Delta$ consists of finitely many elements for $E_n$ with $n < 9$,
and of infinitely many elements if $n \geq 9$. The Cartan matrix of
$E_n$ implies that the inner product on the root space as we have
defined it is of positive definite signature for $n < 9$, contains one
null-direction if $n=9$, and is of Lorentzian signature
(i.e. $(1,n-1)$) for $n > 9$. 

The lattice dual to the root lattice is called the coweight
lattice, which we will denote as $Q_n$. If we define the fundamental
coweights by 
\be
\inp{\alpha_i}{\omega_j} = \delta_{ij}, \qquad \alpha_i \in \Pi
\ee
then we can describe $Q_{n}$ as generated by linear combinations of
the fundamental coweights with coefficients in $\Z$. It is clear that
the coweight lattice contains the coroot lattice $P_n^{\vee}$, as a sublattice.

We will make much use of the Weyl group $W_{n}$ of $E_{n}$. This is
the group generated by the Weyl reflections $w_i$ in the simple roots,
\be
w_i(\beta) = \beta - \inp{\beta}{\alpha_i^{\vee}} \alpha_i
\ee
The Weyl group leaves the inner product invariant
\be \label{winp}
\inp{w(\alpha)}{w(\beta)}=\inp{\alpha}{\beta} \qquad w \in W_{n}
\ee

The Weyl group includes reflections in the non-simple roots.

\subsection{The subalgebra $H_n$}

In the relevant, dimensionally reduced or unreduced theories, the
$E_n$ algebra's are not manifest, but non-linearly realized. The
relevant variables to describe the theory are captured by the coset
$E_n/H_n$, where $H_n$ is a subgroup of $E_n$. Now in the original
formulation of the problem, these cosets appeared in the dimensional
reduction over space-like directions only, of the 11 dimensional
supergravity in space-time signature (1,10). In that case, the groups
$H_n$ are compact.

Recalling that the compact subgroup of $H_n$ is generated by the
generators $e^k_{\alpha} - e^k_{-\alpha}$, it was proposed in
\cite{Englert:2003py} to modify these to take into account non-compact
generators (we will present a more precise definition in the next
subsection). The generators of $H_n$ are then taken to be 
\be \label{Hgen}
T^k_{\alpha} = e^k_{\alpha} - \epsilon_{\alpha} e^k_{-\alpha}
\ee
As in \cite{Keurentjes:2004bv}, the $\epsilon_{\alpha}$ cannot depend
on $k$, as we will review in a moment.

Also in \cite{Keurentjes:2004bv}, it was argued
that in spite of the infinitely big root system of $E_{11}$, it is
actually sufficient to specify the $\epsilon_{\alpha}$ for a basis of
simple roots. The argument does not depend on the fact that the
underlying algebra is $E_{11}$, and we will briefly repeat the
relevant steps here.

The first step in the argument consists of noting that, due to
equation (\ref{herm}), 
\be \label{defeps}
(T^k_{\alpha})^{\dag} = - \epsilon_{\alpha} T^k_{\alpha},
\ee
such that the sign $\epsilon_{\alpha}$ actually encodes the
hermiticity properties of the generator $T^k_{\alpha}$.

Then we note that the generators $T^k_{\alpha \pm \beta}$ actually
appear in the commutator of $T^k_{\alpha}$ and $T^k_{\beta}$. This,
together with the reality of the structure constants of $E_n$,
immediately implies that the hermiticity properties of $T^k_{\alpha
  \pm \beta}$ follow from those of $T^k_{\alpha}$ and $T^k_{\beta}$.

Then we realize that any generator $T^k_{\alpha}$ can be formed by
taking multiple commutators of the $T_{\alpha_i}$, where the
$\alpha_i$ are the simple roots. From this it follows immediately that
we were correct in asserting that specifying $\epsilon_{\alpha_i}$ for
 the simple roots $\alpha_i$, we have fixed and described the
non-compact form of the algebra completely. Moreover, it is easy to
see that the coefficients $\epsilon_{\alpha}$ depend on $\alpha$, but
not on the degeneracy index $k$.

This information is now transferred to a function $f(\alpha)$, that
is related to $\epsilon_{\alpha}$ by
\be
\epsilon_{\alpha} = \exp (i \pi f(\alpha))
\ee
Because of the properties of the commutator, it follows immediately
that
\be
f(\alpha + \beta) = f(\alpha) + f(\beta)
\ee
and hence $f$ is a linear function, taking values in $\Z_2$. Note that
the minus sign in (\ref{defeps}) is crucial in establishing linearity.

We furthermore note that all such functions are described by
\be
f(\alpha) = \sum_i p_i \inp{\alpha}{\omega_i}
\ee
where $\omega_i$ are the fundamental coweights, and the $p_i$ are
coefficients in $\Z_2$. Different choices for the Dynkin basis of the
algebra are related by the elements of the Weyl group, and we should
therefore not distinguish between $f$ and its images generated by Weyl
reflections. The physical interpretation relies on the intersection of
an $A_n$-algebra with the full algebra and this may change under Weyl 
reflections. This is because the non-trivial Weyl reflection (the ones
that cannot be interpreted as permutations of coordinates) correspond
to sequences of T-dualities, as we will review in section \ref{Weyl}. 
We can now proceed as in \cite{Keurentjes:2004bv}, but there is
actually one more notable point. 

In terms of the complexification $(E_n)^{\C}$ of the algebra, we have
\bea
\exp (i \pi f) h \exp (-i \pi f) & = &  h \label{realf1}\\
\exp (i \pi f) e_{\beta} \exp (-i \pi f) & = & \exp( i \pi f(\beta))
e_{\beta} \label{realf2}
\eea
Note that we have to turn to $(E_n)^{\C}$, because $\exp(i \pi f)$ is
not an element of the the split real form as we have defined it in the
above. These relations mean that $\exp i \pi f$, via the adjoint
action, defines an involution of the algebra $(E_n)^{\C}$. Because
$\exp i \pi f$ represents a group element (of the complex group), this
involution is \emph{inner}. Conversely, if an involution is inner, it
is conjugate to an element of the form $\exp (i \pi f)$, which is on
the maximal torus (the exponentiation of the Cartan sub-algebra). The algebraic
characterization them implies that $f \in Q_n$.

\subsection{Involutions and real forms} \label{realf}

The result of the previous section essentially implies that we are
looking for those involutions on the algebra $E_n$ that are inner. In Lie
algebra theory, the study of involutions is tightly connected to the
study of real forms of Lie algebra's, and we will actually borrow some
results from there. 
 
The central object in the study of real forms of
 semi-simple Lie-groups is that of the Cartan involution
 \cite{Helgason}. From the Cartan involution the non-compact real form
 of the group can be  easily reconstructed.

An involutive automorphism $\theta$ is called a Cartan
involution if $-\inp{X}{\theta Y}$ is strictly positive definite for
all algebra generators $X,Y$. An involution has by definition
eigenvalues $\pm 1$ and the realization of the involution can be
chosen such that the Cartan subalgebra is closed under the involution.

In the supergravity literature the Cartan involution is usually chosen
to be realized in the way that is encoded in a Satake diagram (see
\cite{Helgason} for mathematical background, or
\cite{Keurentjes:2002rc} for a physicists account of the theory with
applications to gravity and supergravity). This
way of realizing the involution can be adapted to all possible real
forms, but we will need it here only to realize the so-called split real form
of the algebra. We call the split real form $E_{n(n)}$, as usual, and
realize it by using a Cartan involution acting on
the root space as 
\be \label{invsplit}
\theta(\alpha) = -\alpha.
\ee
The corresponding  generators of the real form of the algebra are
$h \in {\cal H}$, $e_{\alpha}^k + e_{-\alpha}^k$ and $e_{\alpha}^k -
  e_{-\alpha}^k$ with $\alpha \in \Delta$ (we have anticipated our
  application to infinite dimensional algebra's, where
  $\dim(g_{\alpha})$ can be bigger than one and we will need the
  degeneracy index $k$). This implies that all
  generators of $E_n$ can be formed by linear combinations of elements
  $h$, $e_{\pm \alpha}$ with \emph{real} coefficients.  

For the definition of the denominator sub-algebra $H_n$ we require
only those involutions that are inner. These inner involutions are
specified by a function $f$, and we will now work out the realization
of the real form defined by $f$. 

This real form, that we will call ${\cal E}_n$, defined by the involution
exhibited in (\ref{realf1}) and (\ref{realf2}) has as its  generators
\bea
& ih & \forall \ h \in {\cal H} \label{cart} \\ 
& e^k_{\alpha} - \exp (i \pi f(\alpha)) e^k_{-\alpha} \label{ladmin} & \\
& i \left( e^k_{\alpha} + \exp (i \pi f(\alpha)) e^k_{-\alpha} \right)
\label{ladplus} &  
\eea
Obviously, in our conventions the generator (\ref{cart}) is
anti-hermitian. The hermiticity properties of the generators
(\ref{ladmin}) and (\ref{ladplus}) however depend on the value of the
function $\exp i \pi f(\alpha) = \epsilon_{\alpha}$. 

We can now define the group $H_n$ in the following way. We have the
complexified algebra $(E_n)^{\C}$ and have defined two real forms ,
both embedded in $(E_n)^{\C}$. One is the split real form, denoted as
$E_{n(n)}$, and realized as described below (\ref{invsplit}). There is
a second real form, not necessarily equivalent to the split real form,
that we realize as in the equations (\ref{cart}), (\ref{ladmin}) and
(\ref{ladplus}). This second real form we denote by ${\cal E}_n$, and
it will play only an auxiliary (but crucial) role. We can now define
the denominator subalgebra $H_n$ as the intersection
${\cal E}_n \cap E_{n(n)}$. In a diagram:
\be \label{diag}
\ba{cccl}
(E_n)^{\C} & \supset & {\cal E}_n & \\
\cup & & \cup & \\
E_{n(n)} & \supset & H_n & = {\cal E}_n \cap E_{n(n)}
\ea
\ee

With this explicit realization $H_n$ is generated by generators of the
form (\ref{ladmin}), which again are precisely generators of the form
of equation (\ref{Hgen}) that were previously used to define the
denominator subgroup $H_n$ ad hoc. The relevant groups for
supergravity theory are found on the lower line of (\ref{diag}) as the
coset appearing will be $E_{n(n)}/H_n$. The groups at the upper line
are useful for mathematical exploration, and in particular we will
derive a nice result from the relation to the group ${\cal E}_n$.

Some remarks are in order. First, it is obvious that the diagram
(\ref{diag}) can be generalized to other groups than the $E_n$ series,
and these should play a role in time-like compactification of the
theories in \cite{Breitenlohner:1987dg, Cremmer:1999du,
  Keurentjes:2002xc, Keurentjes:2002rc}. 

Second, the discussion here
makes precise the meaning of the phrase ``temporal involution'',
introduced in \cite{Englert:2003py}. This involution is just an
involution on the complex algebra $(E_n)^{\C}$, defining the real form
${\cal E}_n$, although one uses in effect only its restriction to
$H_n$. An important thing to notice is that all (inner) involutions on
$(E_n)^{\C}$ descend to involutions on $H_n$, but the reverse is not
true. Hence, there may exist real forms of $H_n$ that can not appear
in the denominator of $E_{n(n)}/H_n$ in the context of (super-)gravity
theories. This is related to the fact that in all studies, embeddings
of subalgebra's are (sometimes implicitly) assumed to have certain
regularity properties \cite{Julia:1980gr, West:2001as}. An attempt at a
motivation of the necessity of regularity of sub-algebra's in the
context of (super-) gravity, for finite dimensional algebra's can be
found in \cite{Keurentjes:2002xc}. 

A third remark concerns the fact that in the supergravity literature
one occasionally encounters the phrase ``real forms of
supergravities'', referring to the various variant
supergravities. Although there are many links with the theories of
real forms of the underlying algebra's, as the present discussion and
\cite{Keurentjes:2004bv} demonstrate, it is not quite appropriate to
refer to ``real forms of supergravities''. One obvious reason is that
\emph{different} variant supergravities are described by the
\emph{same} real form of the algebra, see \cite{Keurentjes:2004bv} and
the rest of this paper. We prefer therefore the more neutral
``variant'' above ``real form'', when referring to supergravity.

\subsection{The signature of finite dimensional $H_n$}

We can divide the algebra into hermitian generators, generating
 non-compact symmetries, and anti-hermitian generators, generating
 compact symmetries. If ${\cal G}$ is a finite dimensional real form,
 we can define $n(\cal{G})$ to be the number of non-compact generators
  of ${\cal G}$, and  $c(\cal{G})$ the number of compact
  generators. For these finite dimensional algebra's one obviously has
\be
\dim({\cal G}) = n ({\cal G})+ c ({\cal G})
\ee
The signature or character $\sigma({\cal G})$ of ${\cal G}$ is defined
 to be the difference of these two quantities:
\be
\sigma({\cal G}) = n({\cal G}) - c({\cal G}).
\ee
We furthermore will need the rank $r({\cal G})$ which is the
 dimension of the largest completely reducible Abelian sub-algebra one
 can find.

Now we look at the inner involution that defines the real form for ${\cal
  E}_n$, as well as the algebra $H_n$. Some properties of ${\cal E}_n$
  and $H_n$ are easily related.

First, the non-compact elements of ${\cal E}_n$ are in one to one
correspondence with those roots of $E_n$ on which the involution has
value $-1$. Due to linearity and the $\mod \ 2$ property,
\be
\exp(i \pi f(\alpha)) = \exp(i \pi f(-\alpha))
\ee
Therefore positive and negative roots are paired by the involution,
but actually the definition of the generators (\ref{ladmin}) and
(\ref{ladplus}) imply that non-compact generators also always come in
pairs. The Cartan generators commute with the maximal torus, and
hence all correspond to compact generators, as is explicit in
(\ref{cart}). The non compact elements of $H_n = {\cal E}_n \cap
E_{n(n)}$ are in one-to-one correspondence with the \emph{positive}
roots on which the involution has value $-1$; only the generators
(\ref{ladmin}) are contained in the intersection. Hence
\be \label{ncomph}
n(H_n) = \frac{n({\cal E}_n)}{2}
\ee
The number of compact elements of $H_n$ are given by 
\be \label{comph}
c(H_n)= \frac{c({\cal E}_n)-r({\cal E}_n)}{2}
\ee
where we have used that for the split real form, the dimension of
$H_n$ is equal to the number of positive roots of $E_n$, that can be
computed from
\be
\dim(H_n) = \frac{\dim({\cal E}_n) - r({\cal E}_n)}{2}
\ee
Subtracting (\ref{comph}) from (\ref{ncomph}) and using the
definitions, this little algebra reveals
\be \label{sig}
\sigma(H_n) = \frac{\sigma({\cal E}_n) +r ({\cal E}_n)}{2}
\ee
Given a real form of ${\cal E}_n$, defined by an inner involution, with
given signature $\sigma({\cal E}_n)$ (these can be looked up in tables e.g. in
\cite{Helgason}, or for example in \cite{Keurentjes:2002xc}), we can
immediately, and trivially compute the signature for a particular
possibility for $H_n$. In many instances, the fact that we know the
compact form of the algebra $H_n$, together with the signature
$\sigma(H_n)$ determines the real form of $H_n$ completely. In the
computations in the rest of this article there is only a single
exception to this rule, to be discussed in subsection \ref{excep}.

Equation (\ref{sig}) is one of the core results of our paper. It
is easily seen that it generalizes to all instances of oxidation and
dimensional reduction, that yield a coset sigma model on
$G/H$ with $G$ a split real form. We have computed the list of all
possible denominator subgroups $H$ for a given split Lie-group $G$,
and added it as appendix \ref{app} to this paper. It should also be
obvious how it generalizes to non-split $G$, although we have not
found a formula generalizing (\ref{sig}) that is as compact and elegant.
 
\section{Methods of computation} \label{comp}  

In this section we discuss a methods of computation. First however
we recall an argument on why the methods based on Weyl-reflections
should actually give us the correct signatures for time-like
T-dualities.

\subsection{The $E_n$ Weyl group as the ``self-duality group'' for
  type II strings} \label{Weyl}

This section presents a variant on an argument that can be
found in \cite{Obers:1998fb}. We believe it demonstrates the
applicability of our results, even for those cases where we have to
rely on conjectural symmetries.

In \cite{Obers:1998fb} it is shown that the $E_n$ Weyl-group arises as
a combination of geometrical symmetries with T-duality. It is
instructive to reformulate the geometrical symmetries in terms of the
S-duality of the type $IIB$ superstring, as this will clarify the
precise map between those symmetries that map a type II-theory to
itself, to the Weyl group of $E_n$.

Consider a type $II$ theory compactified on an $n$-torus. The
space-time signature of this torus is not relevant in this section.

The $IIB$-theory has an S-duality symmetry acting on the string
coupling $g$ and the radii of the compact directions $R_i$ as:
\be
\ba{lcl}
\ln g & \rightarrow & - \ln g \\
\ln R_i & \rightarrow & \ln{R_i} - \hlf \ln g
\ea
\ee 
There are also T-duality symmetries mapping the $IIA$-theory to the
$IIB$-theory and vice versa. The T-duality in the $j$-direction acts
as
\be
\ba{lcl}
\ln g & \rightarrow & \ln g - \ln{R_i} \\
\ln R_j & \rightarrow & \ln{R_j} - 2 \delta_{ij} (\ln{R_i})
\ea
\ee 

Given these transformations, we define for $IIA$-theory compactified
on an $n$-torus the transformations $W_j$ with $j \leq n$, and $W_0$
for $n \geq 2$:
\bea
W_1 & = & T_1 S T_1 \\
W_{i+1} & = & T_i S T_i T_{i+1} S T_{i+1} T_i S T_i = T_{i+1} S
T_{i+1} T_i S T_i T_{i+1} S T_{i+1} \qquad i > 1\\
W_0 & = & T_1 S T_1 T_2 S T_2 T_1 S T_2
\eea
The transformation $W_1$ is well-known as the transformation that
exchanges the 11$^{th}$ dimension in $M$-theory with the $1$-direction of
the $IIA$-theory. With this in mind it is also easily shown that the
sequence of dualities described by $W_{i+1}$ lead to nothing but a
permutation of the $i^{th}$ and $(i+1)^{th}$ direction. Note also that
the elements of the Weyl group that change the orientation on the root
lattice (and in particular, the reflections) have an odd number of $S$-entries.

These transformations map the $IIA$-string to itself, or to a variant of
itself. It is easily verified that
\be
(W_i)^2 = 1 
\ee
and if $i \neq j$
\be
(W_i W_j)^{n_{ij}} =1 \qquad n_{ij} = \left\{\ba{c@{, \quad}c}
2 & \textrm{if }  a_{ij} a_{ji} =0 \\
3 & \textrm{if }  a_{ij} a_{ji} =1 \\
\ea \right.
\ee
where $a_{ij}$ is an entry in the Cartan matrix $A_n$ of $E_n$. This
means that the group generated by the above transformations is a \emph{Coxeter
  group}, and that moreover this Coxeter group is isomorphic to the
Weyl group of $E_{n}$. This is actually well-known, but there are a
few advantages to formulating the group this way.

First of all, we observe that we can form all the elements
\be \label{twoT}
T_1 T_i =
\left\{
\ba{c@{, \qquad }c}
W_0 W_2 & i=2; \\
W_i T_1 T_{i-1} W_i & i>2. \\
\ea
\right.
\ee
Together with $W_1 = T_1 S T_1$, we can form any element $T_i T_j$,
and $T_i S T_j$. These in turn can be used as building blocks to build
any duality transformation that maps the $IIA$-theory to itself (or
a variant in a different space-time signature, and/or other signs in
front of the form-field terms). This proves that the Weyl group of
$E_{n}$ is isomorphic to the group of ``self-duality''
transformations, and provides a one-to-one map between Weyl-group
elements and duality transformations, by the identification of generators
\be
W_i \leftrightarrow w_i.
\ee

The argument is translated to $IIB$ theory by performing a T-duality
transformation. It is most convenient to use a T-duality in the $1$
direction for this, to obtain as generators
\bea
W_1 & = & S\\
W_2 & = & S T_1 T_2 S T_1 T_2 S \\
W_{i+1} & = & (S T_i T_{i+1})^3 \qquad i>2 \\
W_0 & = & (S T_1 T_2)^3
\eea
Again the combinations of two T-dualities can be built as in equation
(\ref{twoT}); together with the transformation $W_1 = S$ any duality
transformation mapping type $IIB$ theory to itself can be built from
these.

This alternative view on Weyl group symmetries, which are an obvious
consequence of the dimensionally reduced low energy theory, implies
that these symmetries should lift to symmetries of
the full (not-truncated) string- and $M$-theory (if T- and S-dualities
are exact symmetries of these theories). This adds
credibility to the use of Weyl-group symmetries for the groups $E_9$,
$E_{10}$ and $E_{11}$, where the theory with the full symmetry has not
yet been established. 

What is \emph{not} demonstrated by this argument is how the space-time
signature is encoded in the algebra. For this we do explicitly rely
on the non-linear realizations. However, noting that from $E_n$ with
$n \geq 3$ on, our theory runs exactly parallel with established
results, and using that the extension of the Coxeter group to lower
dimensions involves nothing but duality transformations that perform
coordinate permutations, we may argue that combining the previous
arguments with (Lorentz-)covariance forces the realization of the
space-time signature we have established on us.

\subsection{Explicit realizations of the root space of $E_n$}

Consider the space defined by
$n$-tuples forming vectors $p = (p_1, \ldots p_{n}) \in \R^{n}$, with norm
\be
 ||p||^2 = \sum_{i=1}^{n} p_i^2 - \frac{1}{9} \left(\sum_{i=1}^{n}
 p_i \right)^2,
\ee
and inner product $\inp{ \, }{}$ defined by the norm via
\be \label{inp}
\inp{a}{b} = \frac{1}{2} \left( ||a +b||^2 -||a||^2 - ||b||^2
\right)=\sum_{i=1}^{n} a_i b_i - \frac{1}{9} \left(\sum_{i=1}^{n}
 a_i \right)\left(\sum_{i=1}^{n}
 b_i \right).
\ee
With this inner product, the space always has $n-1$ Euclidean
directions, as the $n-1$ dimensional subspace of vectors $x$ defined by
$\sum_{i=1}^{n} x_i =0$ contains only vectors of positive norm. The 1
dimensional orthogonal complement consisting of vectors $x$ of the
form $x_i = \lambda, \lambda \in \R, \forall i$. These have norm 
\be
\lambda^2 n(1- \frac{n}{9});
\ee
This is positive for $n <9$, negative for $n > 9$, and zero for
$n=9$. Consequently, these spaces have the right properties to serve
as root spaces for the $E_n$ algebra's (these
 choices were inspired by the choice of metric on the $E_{10}$ root
 space in \cite{Damour:2002et, Brown:2004jb}).

We realize the simple roots of $E_{n}$ in the above space as
\bea
\alpha_{n-i} & = & e_i - e_{i+1}, \qquad i = 1, \ldots, n-1;\\
\alpha_0 & = & e_{n-2} + e_{n-1} + e_{n}.
\eea
Here $(e_i)_j = \delta_{ij}$, and the reader should notice that with
the inner product (\ref{inp}) these are \emph{not} unit vectors. The
root lattice of $E_{n}$ consists of
linear combinations of the simple roots, with coefficients in
$\Z$. This lattice can be characterized as 
consisting of those vectors $a$ whose components $a_i$ are integers,
and sum up to three-folds $3k$. The integer $k$ counts the occurrences
of the ``exceptional'' root $\alpha_0$, and for roots equals the level
as defined in \cite{West:2002jj, Nicolai:2003fw, Damour:2002cu}.

The roots at level $0$ coincide with the roots of the algebra
$A_{n-1}$. Its real form is $SL(n,\R)$ whose discrete subgroup
$SL(n,\Z)$ is argued to be the symmetry acting on the compactification
torus $T_n$. The intersection of $A_{n-1}$ with $H_n$ provides us with
a real form of $so(n,\C)$ specifying the space-time signature of the
torus $T_n$.

In the basis we have chosen the Weyl reflections $w_i$ in $\alpha_i$,
$i=1,\ldots,n$ permute entries of the row of
$n$ numbers. The action of these is easily taken into account. 
Signature changing dualities (not acting as space-time coordinate
permutations) are given by Weyl reflections in $\alpha_{0}$, and other
roots that have $\alpha_{0}$ exactly once in their expansion. These
are all of the form 
\bd
\beta_{ijk} = e_i + e_j + e_k, \qquad i<j<k
\ed
These conventions are convenient for exploring compactifications of
$M$-theories. The fundamental coweights can be easily computed, except
for $E_9$ where the computation is impossible because of the
null-direction in the root space. This can however be easily
circumvented by embedding the root space of $E_9$ in that of $E_{10}$,
in the obvious way. The coweights are then specified up to an
arbitrary entry expressing the value on the root of $E_{10}$ that is
absent in $E_9$.

To relate to $IIA$-theories, the root that represents the mixing of
the $11^{th}$ dimension with the rest of space-time obtains a
different interpretation. For convenience, we can take it to be node
$1$. Now the global symmetry on the space-time torus is encoded by the
roots $\alpha_2, \ldots ,\alpha_{n}$. 

Unfortunately, these conventions are well-adapted to $M$- and
$IIA$-theory, and therefore slightly inconvenient for
$IIB$-theory. Roughly an analogous choice for $IIB$-theory would be
to use a space $\R^n$ with inner product
\be
\inp{a}{b} = \sum_{i=1}^{n-1} a_i b_i - \frac{1}{8} \left(\sum_{i=1}^{n-1}
 a_i \right)\left(\sum_{i=1}^{n-1}
 b_i \right) + \hlf a_n b_n.
\ee
Again it is easily shown that this inner product is positive definite
for $n < 9$, that it has a single null direction for $n=9$ and has
Lorentzian signature for $n > 9$.

The simple roots of $E_n$ are then realized as
\bea
\alpha_{n-i} & = & e_i - e_{i+1}, \qquad i = 1, \ldots, n-3;\\
\alpha_0 & = & e_{n-2}-e_{n-1}; \\
\alpha_{1} & = & -2e_n;\\
\alpha_{2} & = & e_{n-2} + e_{n-1} + e_n.
\eea
We now identify an $SL(n-1,\R)$ group, with simple roots $\alpha_0, \alpha_3,
\ldots \alpha_{n-1}$. In these conventions the Weyl reflections
$w_0,w_3 \ldots, w_{n-1}$ generate the permutation group on
space-time. The Weyl reflection $w_{1}$ generates $IIB$
S-duality. Signature changing dualities must come from the Weyl
reflection $w_{2}$. The root $\alpha_{2}$ can be suggestively
decomposed as $e_{n-2} + e_{n-1}$ (representing a 2-form) and $e_n$,
signifying that this 2-form is part of a doublet under the $SL(2,\R)$ factor.

These conventions are ``nice'' for computations that stay well inside
$IIB$. For computations comparing $M$- and $IIA$- on the one hand, and
$IIB$ on the other hand, the results must either be translated to the
other conventions, or one must stick to the abstract formulation.

\subsection{Space-time signature; Diagrammatics}

The cases of 11 dimensional supergravity, and the 10 dimensional $IIA$-
and $IIB$-supergravities rely on different $SL(k,\R)$ subalgebra's. It
is straightforward to extract the space-time signature from these
algebra's following an adaptation of methods of \cite{Englert:2003py, Schnakenburg:2003qw, Keurentjes:2004bv}. Essentially, one can count the number of
generators of $SL(k,\R)$ of the form (\ref{Hgen}) or (\ref{ladmin}),
and divide them into compact and non-compact generators (specified by
$f$). This allows an easy determination of the signature of the
$SO(k-p,p)$ subgroup of $SL(k,\R)$, and hence fixes $p$. 

The remaining generators of $E_n$ are related to (reductions of) form
fields. In \cite{Keurentjes:2004bv} we argued that for $M$-theories,
the presence of $\omega_0$ in the function $f$ encodes the sign in
front of the 4-form term in the 11 dimensional Lagrangian. In the $IIA$
interpretation, it can be seen that the root $\alpha_0$ no longer
corresponds to the 3-form potential, but to the 2-form potential that
results from dimensional reduction, with its 3-form field strength. 
The reasoning of
\cite{Keurentjes:2004bv} carries over the sign of the 3-form field
strength term in the $IIA$ Lagrangian. The $IIA$ interpretation also
alters the interpretation of the root $\alpha_1$. In the $M$-theory
picture, $\alpha_1$ is a root of the $SL(n,\R)$ group. In the
$IIA$-picture, it is seen to correspond to the Kaluza-Klein
vector. The presence of the coweight $\omega_1$ in the function $f$
encodes the sign in front of the 2-form field strength term in the
$IIA$-Lagrangian. From the algebra one can easily see that this sign
in unconventional, if and only if the $IIA$ interpretation is related
to an 11-dimensional interpretation by time-like reduction. The
algebra reproduces the same relation that is well-known from the explicit
dimensional reduction procedure.  

For the $IIB$-interpretation we have the roots $\alpha_1$ and $\alpha_2$
that are not participating in the space-time symmetry group
$SL(k,\R)$. The easiest to interpret of these two is $\alpha_1$. Being
completely orthogonal to the roots of $SL(k,\R)$, it should be clear
that this is a root of an $SL(2,\R)$, precisely the $SL(2,\R)$
appearing as a global symmetry in $IIB$ supergravity. The appearance
of the coweight $\omega_1$ in the function $f$ specifying the real
form indicates whether the denominator subgroup appearing under
$SL(2,\R)$ is compact of non-compact, that is whether the
10-dimensional coset is $SL(2,\R)/SO(2)$ or $SL(2,\R)/SO(1,1)$. The
root $\alpha_2$ corresponds to two-forms of $SL(k,\R)$, forming a
doublet under $SL(2,\R)$. Consequently, the coweight $\omega_2$ must
encode information on the sign of the 3-form terms in the
$IIB$-Lagrangian. There is however a slight subtlety: If the
10-dimensional coset is $SL(2,\R)/SO(2)$ both the 3-form terms in the
Lagrangian have the same sign, and which sign is determined by
$\omega_2$. If the coset is $SL(2,\R)/SO(1,1)$ however, the two
3-forms have different signs. Now $\omega_2$ also contributes to the
sign, but in this case its contribution can always be undone by a
field redefinition, or alternatively, an $S$-duality transformation. 

There are of course more fields in the $IIA$ and $IIB$ Lagrangians. As
we will explain in the sections \ref{IIA} and \ref{IIB}, the signs of
these are completely determined in terms of the other signs.

A nice way to visualize the action of the function $f$ is to inscribe
its values on the simple roots on the corresponding nodes of the
Dynkin diagram. For example for $E_5 \cong D_5$ a particular function
$f$ would be encoded by  
\bd
\ba{cccc}
  &   & 1  &   \\  
1 & 1 & 0 & 1 \\
\ea
\ed
A nice feature of such a visualization is that the action of a
fundamental Weyl reflection is easily transcribed onto the
diagram. From 
\be \label{weylac}
\inp{\alpha_j}{w_i(f)} = \inp{w_i(\alpha_j)}{f} =
\inp{\alpha_j}{f} - a_{ij}\inp{\alpha_i}{f},
\ee
where $a_{ij}$ corresponds to an entry from the Cartan matrix $A_n$,
we deduce that the action of the Weyl reflection $w_i$ on the diagram
is described by ``add the value of node $i$ to all nodes that are
connected to it, and reduce modulo 2''\footnote{This prescription
  applies to all simply laced algebra's. The extension to non-simply
  laced algebra's is easily deduced from equation (\ref{weylac})}.

As an example, for $E_4 \cong A_4$, one particular orbit of functions,
defined by successive Weyl reflections (which are duality transformations)
is described by the diagrams
\be
\ba{ccc}
  &   & 0  \\
1 & 0 & 0 \\
\ea
\leftrightarrow 
\ba{ccc}
  &   & 0  \\
1 & 1 & 0 \\
\ea
\leftrightarrow 
\ba{ccc}
  &   & 0 \\
0 & 1 & 1 \\
\ea
\leftrightarrow 
\ba{ccc}
  &   & 1 \\
0 & 0 & 1 \\
\ea
\leftrightarrow 
\ba{ccc}
  &   & 1 \\
0 & 0 & 0 \\
\ea
\ee

All these functions describe the coset $SL(5,\R)/SO(4,1)$. The
interpretations of the various diagrams differ however. Interpreting the diagrams as referring to $M-$theory, the horizontal line
represents the $4-$torus. The first diagram (as well as the following 3) is
easily seen to represent a torus signature $(t,s)$ with $|t-s| =2$, while the
last diagram corresponds to $|t-s| =4$.

Alternatively, interpreting these diagrams as relevant to $IIB$, the
vertical line corresponds to the $3-$torus, whereas the node at the far end
indicates whether we are dealing with the coset $SL(2,\R)/SO(2)$ (if
its inscribed value is 0), or $SL(2,\R)/SO(1,1)$ if its inscribed
value is $1$. We then have the choice between $|t-s| =3$ and
$SL(2,\R)/SO(1,1)$, or $|t-s|=1$ and $SL(2,\R)/SO(2)$.

Note how both examples nicely illustrate how, by Weyl
reflections/duality transformations, we can have signs from the space-time
signature ``running'' into the gauge sector, and vice versa.

For $n$ sufficiently large the number of diagrams is large. The above
procedure is however easily implemented in a simple computer
program. We have used such a program to verify the tables we will give
in the following sections.

Another easy check on our results is to compare with
\cite{Keurentjes:2004bv}. There the full classification was done for
$E_{11}$, which is as far as our computation will go, and all
functions $f$ relevant to compactifications of $M$-theory were
found.

By erasing a suitable node of the Dynkin diagram of $E_{11}$, the
diagram splits into the two diagrams of $E_n$ and
$A_{10-n}$. Inscribing the values of the possible functions $f$ on the
nodes, one easily finds the signature of the transverse space-time
from $A_{10-n}$, whereas the real form of the denominator subgroup and
the torus signature follow from the values of $f$ on $E_n$. 

\section{The duality web for $M$-theories: Finite dimensional groups}
\label{Msmall} 

We are primarily interested in the denominator subgroups that can
appear. In 11 and 10 dimensions, we have the variants of $M$- and
$IIA$-theory which were already completely described in
\cite{Hull:1998ym} (see also \cite{Cremmer:1998em}). We therefore
proceed immediately to 9 dimensions, which is the first time a
semi-simple factor appears in the algebra.  

\subsection{Coset symmetries in 9 dimensions}

In 9 dimensions the global symmetry group has algebra $gl(2,\R) \cong sl(2,\R)
\oplus \R$. This algebra has a simple geometric interpretation, the
$\R$ being related to rescaling the volume of the two dimensions that
were reduced away (the volume of the 2-torus if we are discussing
toroidal compactification), and the $sl(2,\R)$-term related to volume
preserving transformations. 

The simple factor is generated by $sl(2,\R)$, which is a real form of
$A_1$. The algebra $A_1$ has only two real forms, and no outer
automorphisms, so both are suitable to our construction. The
denominator subgroup is a 1 dimensional Abelian group, which we denote 
by $T_1$ \footnote{This notation reflects that an Abelian group is
  called a torus. Unlike the usual meaning of the word in physics, the
  mathematical notion does not require the group to be compact.}. 
The real forms, and the predicted Abelian subgroups are
\be
\ba{c@{ \quad : \quad }l@{ \quad \rightarrow \quad }l@{ \quad : \quad }c}
su(2) & \sigma(A_1)=-3 & \sigma(T_1) =-1 & so(2); \\
sl(2,\R) & \sigma(A_1) =1 & \sigma(T_1) =1 & so(1,1). \\
\ea
\ee
Our group theory predicts that there are actually two possible real
forms for the denominator algebra, with $\sigma = \pm 1$. The
denominator subgroup is an Abelian group, $\sigma=1$ indicates it must
be the compact $so(2) \cong u(1)$, while $\sigma = -1$ reveals the non-compact
$so(1,1)$.

Of course $so(2)$ occurs when reducing over two space-, or two
time-like directions respectively, and $so(1,1)$ when there is one
space- and one time-like direction included \cite{Cremmer:1998em}. We
have only included these here to demonstrate that our techniques also
work fine with the simplest examples.

For the relation to $IIB$-theories, we refer the reader to section \ref{IIB}. 

\subsection{Coset symmetries in 8 dimensions}

In 8 dimensions, the global symmetry group has algebra $sl(3,\R)
\oplus sl(2,\R)$. From 8 and less dimensions the global symmetry algebra's are
semi-simple.

We have discussed the real forms for $sl(2,\R)$ in the previous
subsection. The algebra $sl(3,\R)$ is a real form of $A_2$, which has
3 real forms, known as $sl(3,\R)$, $su(2,1)$ and $su(3)$. The
denominator subalgebra must be a real form of $A_1$. But as $sl(3,\R)$
is generated by a Cartan involution that is outer, it does not match
our criteria (notice that its $\sigma(A_2)=3$ would have resulted in
$\sigma(A_1) = 2$ which is impossible for $A_1$). The remaining real
forms imply the following possibilities for the denominator sub-algebra: 
\be \label{so2orso11}
\ba{c@{ \quad : \quad }l@{ \quad \rightarrow \quad }l@{ \quad : \quad }c}
su(3) & \sigma(A_2)=-8 & \sigma(A_1) =-3 & so(3); \\
su(2,1) & \sigma(A_2) =0 & \sigma(A_1) =1 & so(2,1). \\
\ea
\ee

Combining these with the results of the previous section, one finds
that there are 4 possibilities for the denominator sub-algebra, being
$so(3) \oplus so(2)$, $so(3) \oplus so(1,1)$, $so(2,1) \oplus so(2)$
and $so(2,1) \oplus so(1,1)$. 

The $SL(2,\R)$ factor appears because the (axionic) scalar $\psi$,
formed from reducing the 11-dimensional 3 form over the 3-torus,
combines with the (dilatonic) scalar $\phi$ representing the volume of
this 3-torus, into a realization of the coset $SL(2,\R)/SO(2)$ or
$SL(2,\R)/SO(1,1)$. The difference between the two cosets manifests
itself as a relative sign between the dilatonic and the axionic
scalar. The relevant part of the Lagrangian is 
\be \label{sl2coset}
-\hlf \left( \df{\dif \phi} \wedge \dif \phi \pm e^{2 \phi} \df{\dif \psi} \wedge
\dif \psi \right), 
\ee   
The plus appears for the coset $SL(2,\R)/SO(2)$, while the minus-sign
indicates the coset $SL(2,\R)/SO(1,1)$. Which one of the two is realized
depends on two signs: The sign of the 4-form term in the
11-dimensional theory, and a sign coming from the signature of the
3 dimensions one is reducing over.

The theories in signatures $(1,10)$ $(6,5)$, and $(9,2)$ have a
conventional sign in front of the 4-form kinetic term
\cite{Hull:1998ym}, giving a
positive sign in our computation, whereas the
other ones have an extra minus sign. The sign coming from the 3-torus
is plus if the number of time-dimensions is even; otherwise it is minus.

The sign in equation (\ref{sl2coset}) is given by multiplying these
two signs. The plus, indicating $SO(2)$ as denominator subgroup,
appears if the 11-dimensional 4-form term is conventional and the 3
dimensions include an even number of time directions, or if there is
an unconventional sign in 11-dimensions, combined with an odd number
of time directions. In the other cases, the minus sign appears, and
the denominator subgroup is $SO(1,1)$. 

We summarize our findings in table \ref{tab1}.

\TABLE{
\begin{tabular}{|c|cccc|c|}
\hline
$\R_{m,n} $ & \multicolumn{4}{c|}{$T_{p,q}$} & $h$ \\
\hline
$(0,8)$ & $(1,2)$& $(2,1)$&        &        & $so(2,1) \oplus so(1,1)$ \\
$(1,7)$ & $(0,3)$&        &        &        & $so(3) \oplus so(2)$ \\
$(1,7)$ &        & $(1,2)$&        &        & $so(2,1) \oplus so(2)$ \\
$(2,6)$ &        & $(0,3)$& $(3,0)$&        & $so(3) \oplus so(1,1)$ \\
$(3,5)$ &        &        & $(2,1)$&        & $so(2,1) \oplus so(2)$ \\
$(3,5)$ &        &        &        & $(3,0)$& $so(3) \oplus so(2)$  \\
$(4,4)$ &        &        & $(1,2)$& $(2,1)$& $so(2,1) \oplus so(1,1)$ \\
\hline
\end{tabular}
\caption{Dualities of $M$-theories in 8 dimensions} \label{tab1} }

Table \ref{tab1} is the first of a series of tables that all exhibit
the same structure. In the column under $\R_{m,n}$ we have denoted the
signature of the transverse space-time. Then there are a number of
columns representing tori $T_{p,q}$ of various signatures. The reader
can reconstruct which theory we are dealing with by simply computing
the signature $(m+p,n+q)$ of the overall space-time. Each row in our
table represents a group of theories that can be transformed into one
another by duality symmetries. The last entry in the row is the
denominator sub-algebra $h$, that is common to all entries in the 
row. The various descriptions in each row are related by
dualities. The space-time signatures that are not in this table have $n < m$
in $\R_{m,n}$. The answer for these theories can be found by simply
interchanging $m \leftrightarrow n$, $p \leftrightarrow q$, which
gives all the remaining configurations. 

The groups $SO(3) \times SO(2)$, $SO(2,1) \times SO(1,1)$ and $SO(2,1) \times
SO(2)$ can be found in \cite{Hull:1998br, Cremmer:1998em, Hull:1998vg,
  Duff:2003ec, Hull:2003mf}\footnote{The papers \cite{Hull:1998br,
    Hull:1998vg} contain an unfortunate typo in their answer for
  time-like compactification of conventional 11 dimensional supergravity to 8
  dimensions.}. In addition we find that also the group $SO(3) \times
SO(1,1)$ is possible, and that it is crucial to complete the duality web. 

When compactifying on a 3-torus, one can reach at most one of the
other theories. This is always accompanied with a double sign change:
Both the four-form term, and the signature of the torus are
different. That in all those cases the theories are described by a duality
group that has as second factor $SO(1,1)$ is a coincidence; dropping
the requirement of supersymmetry, and studying other signatures one
sees that $SO(2)$ is not ruled out by principle, but simply does not
occur in the $M$-theory duality chain.  

\subsection{Coset symmetries in 7 dimensions}

For 7 dimensions and below, the space-time symmetries and the 3-form
gauge field merge into a simple global symmetry group. In 7 dimensions
the relevant algebra is $sl(5,\R)\cong A_4$. $A_4$ has 4 inequivalent
real forms, of which 3 are generated by inner automorphisms. The
denominator sub-algebra must be a real form of $B_2$, and doing the
computation leads to the following possibilities:  

\be
\ba{c@{ \quad : \quad }l@{ \quad \rightarrow \quad }l@{ \quad : \quad }c}
so(5) & \sigma(A_4) = -24 & \sigma(B_2) = -10 & so(5);\\
su(4,1) & \sigma(A_4) =-8 & \sigma(B_2) = -2 & so(4,1);\\
su(3,2) & \sigma(A_4) = 0 & \sigma(B_2) = 2 & so(3,2).\\
\ea
\ee

This computation has reproduced the groups found in \cite{Hull:1998br,
  Cremmer:1998em, Hull:1998vg}. 

The symmetries of space-time are embedded in a real form of the
algebra $so(4,\C) \cong A_1 \oplus A_1$. To obtain the space-time
signature, one has to study the embedding of $A_3$ in $A_4$, and the
consequences for the $A_1 \oplus A_1$ embedding in $B_2$. Even without
the details, it is already clear that $so(5) \supset so(4)$, that
$so(4,1) \supset so(4), so(3,1)$ and $so(3,2) \supset so(3,1),
so(2,2)$. The generators that are in $B_2$, but not in $A_1 \oplus
A_1$ correspond to the 4-form sector, and there is a sign difference
for the corresponding scalars if for example the $so(4)$ comes from
$so(4,1)$ or $so(5)$. Tracing this sign back to 11 dimensions we
arrive at table \ref{tab2}.   

\TABLE{
\begin{tabular}{|c|cccc|c|}
\hline
$\R_{m,n} $ & \multicolumn{4}{c|}{$T_{p,q}$} & $h$ \\
\hline
$(0,7)$ & $(1,3)$& $(2,2)$&        &        & $so(3,2)$ \\
$(1,6)$ & $(0,4)$&        &        &        & $so(5)$   \\
$(1,6)$ &        & $(1,3)$& $(4,0)$&        & $so(4,1)$ \\
$(2,5)$ &        & $(0,4)$& $(3,1)$&        & $so(4,1)$\\
$(2,5)$ &        &        &        & $(4,0)$& $so(5)$\\  
$(3,4)$ &        &        & $(2,2)$& $(3,1)$& $so(3,2)$\\
\hline
\end{tabular}
\caption{Dualities of $M$-theories in 7 dimensions} \label{tab2} }

Obviously, the compactification of $M_{(1,10)}$ theory on
a Euclidean 4-torus stands isolated, and corresponds to the symmetry algebra $so(5)$. By symmetry, the same applies to $M_{(10,1)}$ on a
time-like 4-torus. Also for the theory in space-time signature $(6,5)$, compactified on a torus with spatial directions
only the duality algebra is $so(5)$ and a transition to another theory
is not possible. This is due to the fact that the $(6,5)$ theory has
the conventional four-form sign. By symmetry, the same applies to
$(5,6)$-theory on a time-like 4-torus.

For all other combinations of signs, duality transitions between
theories of different signatures are possible. 

\subsection{Coset symmetries in 6 dimensions} \label{excep}

In 6 dimensions, the global symmetry is $Spin(5,5)$.
The denominator subgroups that can occur are given by our standard
computation based on equation (\ref{sig}): 
\be
\ba{c@{ \quad : \quad }l@{ \quad \rightarrow \quad }l@{ \quad : \quad }c}
so(10) & \sigma(D_5) =-45 & \sigma(B_2 \oplus B_2) =-20 & so(5) \oplus
so(5);  \\ 
so(2,8) & \sigma(D_5) = -13 & \sigma(B_2 \oplus B_2) = -4 & so(4,1)
\oplus so(4,1); \\ 
so^*(10) & \sigma(D_5) =-5 & \sigma(B_2 \oplus B_2) =0 & so(5,\C); \\
so(4,6) & \sigma(D_5) = 3 & \sigma(B_2 \oplus B_2) = 4 & so(3,2)
\oplus so(3,2). \\ 
\ea
\ee
The computation of the signature is not decisive in the case when the
signature $\sigma(B_2 \oplus B_2) =0$, at first sight this leaves
$so(4,1) \oplus so(3,2)$ and $so(5,\C)$ as options. Studying the
embedding of the $B_2 \oplus B_2$ algebra in $D_5$, however, it is 
easily seen that there is a symmetry between the two $B_2$-factors,
ruling out the first option and fixing the algebra to be the one of
$so(5,\C)$. 

Although they can be computed directly, also here a fairly intuitive
way of understanding the possible space-time signatures is
possible. Studying how the rotations in the 5 reduced dimensions are
embedded in the $so(5) \oplus so(5)$ of the $(1,10)$ theory reduced on 5
Euclidean dimensions, one sees that it is the diagonal algebra that
represents the symmetries of these dimensions. Correspondingly, the
diagonal algebra in $so(4,1) \oplus so(4,1)$ is obviously $so(4,1)$,
and the one in $so(3,2) \oplus so(3,2)$ is $so(3,2)$, so these
theories must correspond to torus signatures $(4,1)$ and $(1,4)$, and
$(3,2)$ and $(2,3)$ respectively.

For the $so(5,\C)$ algebra the situation is a little more
involved. The embeddings follow from writing the algebra as $so(5)
\oplus i \ so(5)$, where $i$ is the imaginary unit. But as also $so(3,2)
\oplus i \ so(3,2)$ and $so(4,1) \oplus i \ so(4,1)$ result in $so(5,\C)$,
it appears that all signatures are possible, and this is indeed
confirmed in a direct computation. Putting all results together, we
arrive at the following table.
 
\TABLE{
\begin{tabular}{|c|cccc|c|}
\hline
$\R_{m,n} $ & \multicolumn{4}{c|}{$T_{p,q}$} & $h$ \\
\hline
$(0,6)$ & $(1,4)$& $(2,3)$& $(5,0)$&        & $so(5,\C)$ \\
$(1,5)$ & $(0,5)$&        &        &        & $so(5) \oplus so(5)$   \\
$(1,5)$ &        & $(1,4)$& $(4,1)$&        & $so(4,1)\oplus so(4,1)$\\
$(1,5)$ &        &        &        & $(5,0)$& $so(5) \oplus so(5)$\\
$(2,4)$ &        & $(0,5)$& $(3,2)$& $(4,1)$& $so(5,\C)$\\  
$(3,3)$ &        &        & $(2,3)$& $(3,2)$& $so(3,2) \oplus so(3,2)$\\
\hline
\end{tabular}
\caption{Dualities of $M$-theories in 6 dimensions} \label{tab3} }

Notice that in this dimension, for the first time, there is a duality
group that connects 3 $M$-theories. It seems that the algebra $so(3,2)
\oplus so(3,2)$ has not appeared in the literature before. The reader
who studies our table \ref{tab3} will notice that it occurs in a part
of the duality web not studied in detail in \cite{Hull:1998vg,
  Hull:1998ym}, only occurring for the $(6,5)$ and $(5,6)$ theory
compactified on a torus of signature $(3,2)$ or $(2,3)$. There is
therefore no contradiction with earlier results, though the algebra
$so(3,2) \oplus so(3,2)$ is
really necessary to complete the duality web. 

\subsection{Coset symmetries in 5 dimensions}

In 5 dimensions, the global symmetry becomes the exceptional
$E_{6(6)}$. Our usual computation for the possible denominator subgroups gives:
\be
\ba{c@{ \quad : \quad }l@{ \quad \rightarrow \quad }l@{ \quad : \quad }c}
e_{6(-78)} & \sigma(E_6) =-78 & \sigma(C_4) =-36 & sp(4); \\
e_{6(-14)} & \sigma(E_6) =-14 & \sigma(C_4) =-4 & sp(2,2);\\
e_{6(2)} & \sigma(E_6) = 2 & \sigma(C_4) = 4 & sp(4,\R).\\
\ea
\ee

The space-time rotations in the compact group $sp(4)$ are hidden in
its regular subalgebra $u(4) \cong su(4) \oplus u(1)$, where one has
to remember that $su(4) \cong so(6)$. Similar relations are true for
the other real forms of $A_3$, as $su^*(4) \cong so(1,5)$, $su(2,2)
\cong so(4,2)$ while $sl(4,\R) \cong so(3,3)$. It then remains to
identify the subalgebra's: $sp(4) \supset su(4)$, $sp(2,2) \supset
su^*(4), su(2,2)$ (the first of these embeddings can be easily seen
from the Satake diagram of $su^*(4)$), and $sp(4,\R) \supset
sl(4,\R),su(2,2)$. The reader less familiar with these groups may want
to try an explicit computation.

\TABLE{
\begin{tabular}{|c|cccc|c|}
\hline
$\R_{m,n} $ & \multicolumn{4}{c|}{$T_{p,q}$} & $h$ \\
\hline
$(0,5)$ & $(1,5)$& $(2,4)$& $(5,1)$&        & $sp(2,2)$ \\
$(0,5)$ &        &        &        & $(6,0)$& $sp(4)$\\
$(1,4)$ & $(0,6)$&        &        &        & $sp(4)$   \\
$(1,4)$ &        & $(1,5)$& $(4,2)$& $(5,1)$& $sp(2,2)$ \\
$(2,3)$ &        & $(0,6)$& $(3,3)$& $(4,2)$& $sp(4,\R)$\\  
\hline
\end{tabular}
\caption{Dualities of $M$-theories in 5 dimensions} \label{tab4}}

Apart from the groups mentioned previously in the literature, we also
find the group $Sp(4,\R)$. Table \ref{tab4} reveals its place in the
duality web. The group $Sp(4,\R)$ appears in a part of the duality web
not explored in detail in \cite{Hull:1998vg, Hull:1998ym}, so there is
no contradiction with earlier results. Note furthermore that again the
$(6,5)$ and $(5,6)$ theories allow a compact denominator group for a
6-torus with space- or time-like dimensions only. This is the last
dimension for which this happens, for a 7-torus the signature must be
mixed for $(6,5)$ and $(5,6)$ theories.

\subsection{Coset symmetries in 4 dimensions}

In 4 dimensions, the global symmetry of the theory is $E_{7(7)}$. The
algebra $E_7$ has no outer automorphisms, and as a matter of fact none
of the $E_n$ algebra's with $n \geq 7$ has. Hence we find a one to one
correspondence between possible real forms of $E_{7(7)}$, and the possible
real forms of the algebra $A_7$ that can appear in the denominator
sub-algebra. Our computation gives the following possibilities: 
\be
\ba{c@{ \quad : \quad }l@{ \quad \rightarrow \quad }l@{ \quad : \quad }c}
e_{7(-133)} & \sigma(E_7) =-133 & \sigma(A_7) =-63 & su(8);  \\
e_{7(-25)} & \sigma(E_7) =-25 & \sigma(A_7) =-9 & su^*(8); \\
e_{7(-5)} & \sigma(E_7) = -5 & \sigma(A_7) = 1 & su(4,4); \\
e_{7(7)} & \sigma(E_7) = 7 & \sigma(A_7) = 7 & sl(8,\R). \\
\ea
\ee

These organize in the duality web according to table \ref{tab5}.

\TABLE{
\begin{tabular}{|c|ccccc|c|}
\hline
$\R_{m,n} $ & \multicolumn{5}{c|}{$T_{p,q}$} & $h$ \\
\hline
$(0,4)$ & $(1,6)$& $(2,5)$& $(5,2)$& $(6,1)$&        & $su^*(8)$ \\
$(1,3)$ & $(0,7)$&        &        &        &        & $su(8)$   \\
$(1,3)$ &        & $(1,6)$& $(4,3)$& $(5,2)$&        & $su(4,4)$ \\
$(2,2)$ &        & $(0,7)$& $(3,4)$& $(4,3)$& $(7,0)$& $sl(8,\R)$\\  
\hline
\end{tabular}
\caption{Dualities of $M$-theories in 4 dimensions} \label{tab5} }

We have again found a possible denominator group, $SL(8,\R)$, that has
not appeared in the literature before, but again there is no
contradiction with earlier results. 

\subsection{Coset symmetries in 3 dimensions}

In 3 dimensions the global symmetry group is $E_{8(8)}$. The algebra
$E_8$ can appear in 3 real forms, which lead to the following
possibilities for denominator subgroups. 

\be
\ba{c@{ \quad : \quad }l@{ \quad \rightarrow \quad }l@{ \quad : \quad }c}
e_{8(-248)} & \sigma(E_8) =-248 & \sigma(D_8) =-120 & so(16);  \\
e_{8(-24)} & \sigma(E_8) =-24 & \sigma(D_8) =-8 & so^*(16); \\
e_{8(8)} & \sigma(E_8) = -8 & \sigma(D_8) = 8 & so(8,8). \\
\ea
\ee
The duality web is reproduced in table \ref{tab6}.

\TABLE{
\begin{tabular}{|c|ccccc|c|}
\hline
$\R_{m,n} $ & \multicolumn{5}{c|}{$T_{p,q}$} & $h$ \\
\hline
$(0,3)$ & $(1,7)$& $(2,6)$& $(5,3)$& $(6,2)$&        & $so^*(16)$\\
$(1,2)$ & $(0,8)$&        &        &        &        & $so(16)$\\
$(1,2)$ &        & $(1,7)$& $(4,4)$& $(5,3)$& $(8,0)$& $so(8,8)$\\
\hline
\end{tabular}
\caption{Dualities of $M$-theories in 3 dimensions} \label{tab6} }

This time we only find groups encountered previously in the
literature. All $M$-theories and all possible signatures for the
$8$-torus are described by this small set of groups.

\section{Low dimensions: Infinite dimensional groups}\label{Mlarge}

Reducing to less than 3 dimensions, we encounter infinite
dimensional groups. For the application we are discussing here, it
does not really matter whether we are discussing dimensionally reduced
theories \cite{Julia:1982gx}, or conjectured formulations of the full,
unreduced theory with a hidden symmetry (such as the proposals of
\cite{Damour:2002cu, West:2001as}) , essentially because
the cosets refer to the zero-mode spectrum only. As the real forms of
the infinite groups we encounter are unfamiliar, we devote some
discussion to explicit realizations. 

\subsection{2 dimensions: real forms of $E_9$ and $H_9$}

In two dimensions we are dealing with the coset $E_{9(9)}/H_9$. There
are several new features, the most significant one of course being the
fact that we are dealing with infinite-dimensional groups
here. Furthermore we will find a group that according to our previous
criteria can appear as a denominator subgroup, but detailed
computation reveals that it does not appear in the $M$-theory duality
web, because it can only correspond to space-time signatures incompatible with
supersymmetry. 

The denominator subgroup $H_9$ can appear in 4 possible real forms,
that we will denote by $H_9^{c}, H_9^{n1}, H_9^{n2}$ and
$H_9^{n3}$. The group $H_9^c$ is the compact real form, and has been
denoted in the past as $K(E_9)$ and $SO(16)^{\infty}$
\cite{Julia:1980gr, Nicolai:kz, Brown:2004jb}. The groups $H_9^{n1},
H_9^{n2}$ and $H_9^{n3}$ are non-compact real forms. We should not by
analogy to the compact case denote these by $SO(8,8)^{\infty}$ or
$SO^*(16)^{\infty}$ or similar; such a notation would be ambiguous, as
for example $SO^*(16)$ is a subgroup of both $H_9^{n1}$ and
$H_9^{n2}$, and on the other hand $H_9^{n2}$ has $SO(8,8)$ as well as
$SO^*(16)$ as subgroups. The reader can verify this by direct
computation, or by using our tables and decompactifying time and
space-like dimensions in the theories with these groups. Which real
form of $H_9^{\C}$ is obtained depends on how one constructs the group
from the infinite tower of $D_8$ representations. Note that the notion
of signature of the algebra, which distinguishes the finite
dimensional real forms is ill-defined for these infinite dimensional groups 

Only the groups $H_9^c$, $H_9^{n1}$ and $H_9^{n2}$ appear in the
duality web, that is reflected in table \ref{tab7}. 
\TABLE{
\begin{tabular}{|c|ccccc|c|}
\hline
$\R_{m,n} $ & \multicolumn{5}{c|}{$T_{p,q}$} & $h$ \\
\hline
$(0,2)$ & $(1,8)$& $(2,7)$& $(5,4)$& $(6,3)$& $(9,0)$& $H_{9}^{n1}$\\
$(1,1)$ & $(0,9)$&        &        &        &        & $H_9^c$\\
$(1,1)$ &        & $(1,8)$& $(4,5)$& $(5,4)$& $(8,1)$& $H_9^{n2}$\\
\hline
\end{tabular}
\caption{Dualities of $M$-theories in 2 dimensions} \label{tab7} }

The fourth real form $H_9^{n3}$ does not appear in
 this table. It is associated to combinations of space-time
 signatures, and signs for the gauge fields that cannot appear in
 compactifications of theories that descend from an 11 dimensional
 supersymmetric theory. 

A particular way to construct $H_9^{n1}$ is as follows: First we make an
$E_9$ level decomposition, by decomposing with respect to the
horizontal $A_8 = SL(9,\R)$ algebra. When we project to $H_9$, we turn all
generators that are composed of ladder operators at levels $\pm k$
where $k$ is odd, into non-compact ones, and the remaining ones into
compact ones. It is easily seen that the algebra obtained this way closes.

For $H_9^{n2}$ we make the $E_9$ level decomposition that is more
common in the mathematical literature, with respect to $E_8$, and then
decompose to $SO(16)$. We have an infinite tower of repeating irreps,
that are either the $\mathbf{120}$ or the spin irrep
$\mathbf{128}$. Projecting to $H_9$ all $\mathbf{120}$ irreps are
paired, except for the one at level $0$. We pair the $\mathbf{128}$ at
level $k$ with the one at level $-k$, the $\mathbf{128}$ at level 0 is
paired with itself. The $\mathbf{120}$ irreps are then projected to
compact generators while all the $\mathbf{128}$ correspond to
non-compact ones. This is the real form of $H_9^{n2}$ that corresponds
to the split real form of $E_{9(9)}$. 

For $H_9^{n3}$ we make an $E_9$ level decomposition in $E_8$ irreps,
and then decompose these to $E_7 \oplus su(2)$. Under this
decomposition $\mathbf{248} \rightarrow \mathbf{(133,1) \oplus (1,3)
  \oplus (56,2)}$. When projecting to $H_9$ we turn all generators of
$\mathbf{(56,2)}$ into non-compact ones, and the remaining ones to
compact ones. The corresponding real form of $E_9$ can be
characterized as the affine Lie algebra built on the real form
$E_{8(-24)}$ of $E_8$. 

The coset $E_{8(-24)}/(E_7 \times SU(2))$ can appear in a 3
dimensional coset theory \cite{Gunaydin:1983rk}. This theory allows a
supersymmetric extension (with at most 8 supersymmetries), and can in
turn be oxidized to a 6 dimensional theory
\cite{Keurentjes:2002rc}. It should be expected that the real form of
$E_9$ implied by $H_9^{n3}$ will appear in the compactification of
this theory to 2 dimensions.   
 
\subsection{1 dimension: real forms of $E_{10}$ and $H_{10}$}

For $E_{10(10)}$ there are three possible real forms. In the theories
that descend from an 11 dimensional supersymmetric theory, only the
compact form, and one of the two real forms can occur.

We have summarized the results of the computation in table
\ref{tab8}, although it has almost trivial content.
 
\TABLE{
\begin{tabular}{|c|cccccc|c|}
\hline
$\R_{m,n} $ & \multicolumn{6}{c|}{$T_{p,q}$} & $h$ \\
\hline
$(1,0)$ &    & $(1,9)$& $(4,6)$& $(5,5)$& $(8,2)$& $(9,1)$& $H_{9}^{n1}$\\
$(1,0)$ & $(0,10)$&        &        &        &        &   & $H_9^c$\\
\hline
\end{tabular}
\caption{Dualities of $M$-theories in 1 dimensions} \label{tab8} }

The compact real form $H_{10}^c$ represents the symmetries of the
theories in space-time signatures
$(1,10)$ and $(10,1)$, where the single time, resp. space dimension is
kept transverse, and the other ones are compactified. The non-compact
form we have called $H_{10}^{n1}$ describes all the other situations
arising from compactification of 11 dimensional supergravity theories.
Another non-compact form $H_{10}^{n2}$ describes theories
in space-time signatures, or with signs in front of the 4-form
gauge-field terms that cannot occur in the $M$-theory duality web. 

A particular way to construct $H_{10}^{n1}$ is as follows: Decompose
$E_{10}$ with respect to the leftmost node into $D_9 \cong so(18)$
irreps. The result is an infinite towers of irreps of which some are
congruent to either of the two spin-irreps\footnote{For the reader
  unfamiliar with the concept of congruent irreps: this means that the
  weights of these  irreps correspond to a weight of a spin irrep,
  plus an element of the root lattice.}, and some are congruent to the
vector or adjoint irreps. We then project to $H_{10}$, by setting all
generators that correspond to irreps congruent to spin irreps of
$so(18)$ to non-compact generators, and the rest to compact ones.
 
The real form $H_{10}^{n2}$, that cannot occur in the $M$-theory duality
web can be constructed as follows. We decompose with respect to the
exceptional node, into $SL(10)$ irreps. We then project all generators
at odd levels (where now we define level with respect to $SL(10)$) to
non-compact generators, and all the generators at even levels to compact ones.

\subsection{0 dimensions: real forms of $E_{11}$ and $H_{11}$}

The essential computations for this case have been done in
\cite{Keurentjes:2004bv}, where they where elaborated upon in great
detail. We therefore only repeat the conclusions of this paper: There
are 4 possible denominator subgroups. Of these, only one corresponds
to the signs appropriate for $M$-theory and its cousins, the $M^*$ and
$M'$ theories. Moreover, it can be demonstrated that this choice of
signs allows all the 11 dimensional supergravity theories, but no others.

It may be worth noting that, with $n \geq 3$, for $E_n$ with $n$ odd
we have always found 4 possible real forms, whereas for $E_n$ with $n$
there appear to be always 3 real forms ( are restricting to
real forms generated by inner involutions). In spite of the
empirical truth of this assertion (we have checked it also for some
cases of $E_n$ with $n > 11$), we have not managed to find a
simple proof of it. 

\section{The duality web for $IIA$ theories} \label{IIA}

All variant $IIA$-theories can be found by suitable compactification
of $M$-theories on either a time- or a space-like circle. The
relations are \cite{Cremmer:1998em, Hull:1998vg, Hull:1998ym} 
\be
\ba{cccccccc}
\multicolumn{2}{c}{ M_{(1,10)}} && \multicolumn{2}{c}{ M_{(2,9)}} &&
\multicolumn{2}{c}{ M_{(5,6)}} \\
\multicolumn{2}{c}{s \swarrow \searrow t} && \multicolumn{2}{c}{s \swarrow
  \searrow t} && \multicolumn{2}{c}{s \swarrow \searrow t} \\
IIA_{(1,9)} & IIA_{(0,10)}&& IIA_{(2,8)} & IIA^*_{(1,9)}&&
IIA_{(5,5)} & IIA_{(4,6)} \\
\ea
\ee
\be
\ba{cccccccc}
\multicolumn{2}{c}{ M_{(10,1)}} && \multicolumn{2}{c}{ M_{(9,2)}} &&
\multicolumn{2}{c}{ M_{(6,5)}} \\
\multicolumn{2}{c}{s \swarrow \searrow t} && \multicolumn{2}{c}{s \swarrow
  \searrow t} && \multicolumn{2}{c}{s \swarrow \searrow t} \\
IIA_{(10,0)} & IIA_{(9,1)}&& IIA^*_{(9,1)} & IIA_{(8,2)}&&
IIA_{(6,4)} & IIA^*_{(5,5)} \\
\ea
\ee
Here $s$ signifies compactification on a space-like circle, whereas
$t$ stands for compactification on a time-like circle. It follows that
the duality groups for the resulting variants of $IIA$-theory can be
immediately deduced from their $M$-theory ancestors, as a $IIA$-theory
in space-time signature $(p,q)$ has the same dualities as the
$M$-theory in signature $(p+1,q)$ if the theories are related by
compactification on a time-like circle, whereas it has the same
dualities as $M$-theory in signature $(p,q+1)$ if they are related by
compactification on a space-like circle. In particular, table 3 from
\cite{Hull:1998vg} is reproduced by collecting from our tables the
entries for compactification of $M_{(2,9)}$-theory, onto a torus with
1-time direction, leaving 1 other time-direction for the transverse
space.  

A useful remark in this context is that the fact that the $IIA$-theories
derive from compactification of $M$-theories implies that there are
relations between the signs of various terms. Alternatively, these
signs are reflected in the algebra, and can also be derived from
this perspective.

In particular the $B_{(2)}$-field 2-form, and the $C_{(3)}$-form in
$IIA$ have the same 11 dimensional origin. The sign of kinetic term of the
10-dimensional $C_{(3)}$-form is inherited from its 11 dimensional
ancestor. The $B_{(2)}$-term picks up an extra sign if one reduces over a
time-like direction. The sign of the Kaluza-Klein vector term
also has an unconventional sign precisely for time-like
compactification. Consequently, one always has the identity
\be \label{eqsigns}
\textrm{sign}(C_{(1)}) \cdot \textrm{sign}(C_{(3)}) = \textrm{sign}(B_{(2)})
\ee
Comparing with table 1 in \cite{Hull:1998ym} this is easily
verified. Actually this relation extends beyond supergravity. Having a
theory with the same field content as the bosonic sector of $IIA$ theory
it can only be oxidized to 11 dimensions, and it will have only have
the proper symmetric space structure in lower dimensions if equation
(\ref{eqsigns}) is obeyed.

We see that the real form of the $IIA$ theory is completely specified
by: the space-time signature; the sign of $C_{(1)}$ denoted as
$\sigma_1$; and the sign of $C_{(3)}$, denoted as $\sigma_3$.

Collecting these in the generalized signature $(t,s, \sigma_1,
\sigma_3)$, we note 
\be
(t,s, \sigma_1, \sigma_3) = (s,t,-\sigma_1,-\sigma_3)
\ee 
Note that the generalized signature of $IIA$ carries just as much
information as the generalized signature for the 11 dimensional
theory (see \cite{Keurentjes:2004bv}). The sign of the ``missing''
dimension is encoded in $\sigma_1$. 
 
\section{The duality web for $IIB$-theories} \label{IIB}

There are two kinds of $IIB$ theories. The first kind has two
ten-dimensional scalars parameterizing the coset $SL(2,\R)/SO(2)$. For
the second kind the two scalars parameterize the
coset $SL(2,\R)/SO(1,1)$. In Hull's notation \cite{Hull:1998vg,
  Hull:1998ym} these are denoted as $IIB^*$ and $IIB'$. The relevant
cosets are captured by the part of the Lagrangian exhibited in
equation (\ref{so2orso11}). Another easy way to distinguish them, from
the bosonic perspective, is to look at the two 3-form field strengths of the 10
dimensional theory, that form a doublet under the $SL(2,\R)$ global
symmetry. As the two fields also form a doublet under the denominator
subgroup, they will appear in the quadratic combination 
\be \label{threeform}
-\hlf \left( e^{\phi} \df{H_{(3)}} \wedge H_{(3)} \pm e^{-\phi} \df{G_{(3)}}
\wedge G_{(3)} \right) 
\ee
The $+$ sign appears for $SO(2)$, the minus sign for $SO(1,1)$. The
Weyl reflection in the single positive root of $SL(2,\R)$ sends $\phi
\rightarrow -\phi$, and $H_{(3)} \leftrightarrow G_{(3)}$, and
therefore changes the overall sign in the $SO(1,1)$ case. This is the
difference between the $IIB^*$ and $IIB'$ theories
\cite{Hull:1998ym}. From the supergravity perspective the distinction
between the two three-forms is arbitrary. In particular,
compactification of either of the two theories leads to the same low
energy theory. 

As in the $IIA$-theory there are relations between the possible
signs. We cannot appeal to a higher dimensional origin of various
signs, but it is easily demonstrated that the algebra leads to
relations.

The sign of the axion $C_{(0)}$ is plus if the coset denominator group
is $SO(2)$ and minus if the group is $SO(1,1)$. The two 2-form
potentials $B_{(2)}$ and $C_{(2)}$ appear as a doublet under the
Abelian group. The quadratic combination appearing in the action has a
relative minus sign between the two components if the group is
$SO(1,1)$ (see equation (\ref{threeform})).

The 4-form $C_{(4)}$ arises in the algebra in the commutator of the
two 2-forms. Putting all these signs together one should have
\be
\textrm{sign}(C_{(0)}) = \textrm{sign}(B_{(2)}) \cdot \textrm{sign}
(C_{(2)}) = \textrm{sign}(C_{(4)}) 
\ee 
Again, with table 2 from reference \cite{Hull:1998ym}, this is easily
verified. It is therefore sufficient to specify the signs of the two 2-form
terms.

We encode everything in a generalized signature $(t,s,\sigma_2,
\sigma_2')$, where we denote by $\sigma_2, \sigma_2'$ the signs of the
2-form terms respectively. If $\sigma_2 \neq \sigma_2'$ one can
interchange the two by an $S$-duality transformation (= Weyl reflection $w_1$).

We now have the equality
\be
(t,s, \sigma_2, \sigma_2') = (s,t,\sigma_2, \sigma_2')
\ee
There is no sign change in the forms ! This can also be seen by noting
that the number of two-forms running over an even number of time-like
directions is 
\be
{t \choose 0}{s \choose 2} + {s \choose 0}{t \choose 2}
\ee
whereas the number of 2-forms running over an odd number of time
directions is
\be
{t \choose 1}{s \choose 1}
\ee
Both these formula's are invariant under $t \leftrightarrow s$
(Compare with the analogous discussion on the 3-form in
\cite{Keurentjes:2004bv}). 

\subsection{Tables for $IIB$-theories}

In this section we will list tables representing groups and dualities
for all toroidal compactifications of $IIB$-theories. The tables will
have the by now familiar structure, apart from one new ingredient.

For the 11 dimensional $M$-theories, we have specified the space-time
signature. There is also an adjustable sign for the 4-form term, but
within the set of supersymmetric $M$-theories, this sign is completely
determined by the space-time signature. This is not so for the
$IIB$-theories, and hence we have indicated in our tables below also the
two signs $\sigma_2, \sigma_2'$, above the columns. The connection
with Hull's notation \cite{Hull:1998vg, Hull:1998ym} is simple: if
$\sigma_2 = \sigma_2'$ the theory corresponds to a $IIB$-theory
without prime or star, whereas if $\sigma_2 \neq \sigma_2'$  the
theory is $IIB'$/$IIB^*$. 

As the $IIB'$- and $IIB^*$-theories are related by a
simple field redefinition (which neither the low-energy theory nor our
algebra can distinguish), they will be collected under a single entry.

We will not list the $10$ and $9$-dimensional theories. The 10
dimensional theories are described in \cite{Hull:1998vg,
  Hull:1998ym}. For the 9 dimensional theories there are some signs
for the form terms, depending on whether one chooses to compactify on
a space- or a time-like direction. These are straightforward to work
out, and the coset symmetry in 9 dimensions is the same as in
10 dimensions \cite{Cremmer:1998em}. 

\TABLE{
\begin{tabular}{|c|ccccc|c|}
\hline
$\R_{m,n} $ & \multicolumn{5}{c|}{$T_{p,q}$} & $h$ \\
\cline{2-6}
            & $++$ &    $+-$  &  $--$    & $+-$    & $--$ & \\
\hline
$(0,8)$ & $(1,1)$& $(1,1)$&        &        &      & $so(2,1) \oplus so(1,1)$\\
$(1,7)$ & $(0,2)$&        &        &        &      & $so(3)   \oplus so(2)$\\
$(1,7)$ &        & $(0,2)$& $(2,0)$&        &      & $so(2,1) \oplus so(2)$\\
$(2,6)$ &        &        & $(1,1)$&        &      & $so(3)   \oplus so(1,1)$\\
$(3,5)$ &        &        & $(0,2)$& $(2,0)$&      & $so(2,1) \oplus
so(2)$\\
$(3,5)$ &        &        &        &        & $(0,2)$& $so(3) \oplus so(2)$\\
$(4,4)$ &        &        &        & $(1,1)$& $(1,1)$& $so(2,1) \oplus
so(1,1)$\\ 
\hline
\end{tabular}
\caption{Dualities of $IIB$-theories in 8 dimensions} \label{tab9} }

The subalgebra's of $sl(3,\R) \oplus sl(2,\R)$, relevant to 8
dimensional theories are collected in \ref{tab9}. In 8 dimensions all
$IIB$-theories living in the same space-time signature, but with
different signs for the forms become T-dual under compactification on
a 2-torus of suitable signature. Of course this is a consequence of
the fact that having 2 directions at our disposal, we can link either
one of them to an intermediate $IIA$-theory. 

Note that a transition to another theory is possible if and only if
the duality algebra contains a $so(2,1)$-term; this is due to the fact
that ``adjacent'' theories have an $so(1,1)$ resp. $so(2)$ local symmetry,
that both have to be contained in the real form of $A_1$ that is
relevant to compactification to 8 dimensions, and hence it must be $so(2,1)$.

The reader may also take notice of the compactifications of the $IIB$
theory in signature $(5,5)$, with conventional signs for the $3$-form
terms, that will give compact symmetry groups for suitable
compactifications up to 5 dimensions, just like the $M$-theories in
signatures $(5,6)$ and $(6,5)$.

\TABLE{
\begin{tabular}{|c|ccccc|c|}
\hline
$\R_{m,n} $ & \multicolumn{5}{c|}{$T_{p,q}$} & $h$ \\
\cline{2-6}
        & $++$   &  $+-$  & $--$   & $+-$   & $++$   &   \\
\hline
$(0,7)$ & $(1,2)$& $(1,2)$& $(3,0)$&         &        & $so(3,2)$ \\
$(1,6)$ & $(0,3)$&        &        &         &        & $so(5)$   \\
$(1,6)$ &        & $(0,3)$& $(2,1)$&         &        & $so(4,1)$ \\
$(2,5)$ &        &        & $(1,2)$& $(3,0)$ &        & $so(4,1)$\\
$(2,5)$ &        &        &        &         & $(3,0)$& $so(5)$ \\
$(3,4)$ &        &        & $(0,3)$& $(2,1)$ & $(2,1)$& $so(3,2)$\\
\hline
\end{tabular}
\caption{Dualities of $IIB$-theories in 7 dimensions} \label{tab10} }

In 7 dimensions transitions between 3 theories are possible if the
subalgebra of $sl(5,\R)$ relevant for the coset is $so(3,2)$. Note how
its decompositions into $so(2,1) \oplus so(2)$, $so(2,1) \oplus
so(1,1)$ and $so(3) \oplus so(2)$ reveal the signatures of the 3-tori
and the 10 dimensional symmetry-group, and that in particular $so(3)
\oplus so(1,1)$ is impossible. Similarly, $so(4,1)$ can only be
decomposed in $so(3) \oplus so(1,1)$ and $so(2,1) \oplus
so(2)$. Putting these facts together inevitably leads to table
\ref{tab10}, that is confirmed by more sophisticated computation.

\TABLE{
\begin{tabular}{|c|cccccc|c|}
\hline
$\R_{m,n} $ & \multicolumn{6}{c|}{$T_{p,q}$} & $h$ \\
\cline{2-7}
        & $++$  & $+-$   & $--$   & $+-$   & $++$ & $--$ & \\
\hline
$(0,6)$ & $(1,3)$& $(1,3)$& $(3,1)$&        &         &         &$so(5,\C)$ \\
$(1,5)$ & $(0,4)$&        &        &        &         &         &$so(5) \oplus so(5)$   \\
$(1,5)$ &        & $(0,4)$& $(2,2)$& $(0,4)$&         &         &$so(4,1)\oplus so(4,1)$\\
$(1,5)$ &        &        &        &        & $(0,4)$ &         &$so(5) \oplus so(5)$\\
$(2,4)$ &        &        & $(1,3)$& $(3,1)$& $(3,1)$ &         &$so(5,\C)$\\  
$(3,3)$ &        &        & $(0,4)$& $(2,2)$& $(2,2)$ & $(4,0)$ & $so(3,2) \oplus so(3,2)$\\ 
\hline
\end{tabular}
\caption{Dualities of $IIB$-theories in 6 dimensions} \label{tab11} }

Table \ref{tab11} collects the results of our computations for
compactifications to 6 dimensions, where the global symmetry algebra
is $so(5,5)$. For a more intuitive understanding of table \ref{tab11},
recall how the $so(4)$ symmetry of 4 compact directions is retrieved
from $so(5) \oplus so(5)$. It is useful to proceed in two steps with
the successive decompositions 
\be
\ba{rcl}
so(5) \oplus so(5) & \rightarrow & so(3) \oplus so(2) \oplus so(3) \oplus
so(2) \\
& \rightarrow & so(3) \oplus so(3) \oplus so(2) \cong so(4) \oplus
so(2), 
\ea
\ee
where on the second line, we have formed the diagonal algebra of the
two $so(2)$ terms, and realized that $so(4)$ is not simple but
consists of two $so(3)$ terms.

With the Lie algebra-isomorphisms $so(2,2) = so(2,1) \oplus so(2,1)$,
and $so(3,\C) = sl(2,\C) = so(1,3)$, and the above embedding, the
reader should be able to reconstruct table \ref{tab11}.

\TABLE{
\begin{tabular}{|c|cccccc|c|}
\hline
$\R_{m,n} $ & \multicolumn{6}{c|}{$T_{p,q}$} & $h$ \\
\cline{2-7}
 & $++$ & $+-$ & $--$ & $+-$ & $++$ & $--$ & \\
\hline
$(0,5)$ & $(1,4)$& $(1,4)$& $(3,2)$& $(5,0)$ &         && $sp(2,2)$ \\
$(0,5)$ &        &        &        &         & $(5,0)$ && $sp(4)$\\
$(1,4)$ & $(0,5)$&        &        &         &         && $sp(4)$   \\
$(1,4)$ &        & $(0,5)$& $(2,3)$& $(4,1)$ & $(4,1)$ && $sp(2,2)$ \\
$(2,3)$ &        &        & $(1,4)$& $(3,2)$ & $(3,2)$ & $(5,0)$ & $sp(4,\R)$\\  
\hline
\end{tabular}
\caption{Dualities of $IIB$-theories in 5 dimensions} \label{tab12}}

The relevant group theory for compactification of $IIB$-theory to
5 dimensions is not difficult. The global symmetry is $E_{6(6)}$. The group
mixing 5 dimensions has algebra $C_2$, and has to embedded in the
denominator algebra which is a real form of $C_4$. It is still
feasible to  realize these algebra's as matrix algebra's. To obtain
the connection with the space-time signature, the isomorphisms $sp(2)
\cong so(5)$, $sp(1,1) \cong so(4,1)$, and $sp(2,\R) = so(3,2)$ for
the real forms of $C_2$ are useful. The reader less familiar with
these algebra's can easily compute the signature and compare the real
form with \cite{Helgason}, \cite{Keurentjes:2002rc}, or another source
listing real forms of the relevant algebra's
 
\TABLE{
\begin{tabular}{|c|cccccc|c|}
\hline
$\R_{m,n} $ & \multicolumn{6}{c|}{$T_{p,q}$} & $h$ \\
\cline{2-7}
 & $++$ & $+-$ & $--$ & $+-$ & $++$ & $--$ & \\
\hline
$(0,4)$ & $(1,5)$& $(1,5)$& $(3,3)$& $(5,1)$& $(5,1)$ && $su^*(8)$ \\
$(1,3)$ & $(0,6)$&        &        &        &         && $su(8)$   \\
$(1,3)$ &        & $(0,6)$& $(2,4)$& $(4,2)$& $(4,2)$ & $(0,6)$ & $su(4,4)$ \\
$(2,2)$ &        &        & $(1,5)$& $(3,3)$& $(3,3)$ & $(1,5)$& $sl(8,\R)$\\  
\hline
\end{tabular}
\caption{Dualities of $IIB$-theories in 4 dimensions} \label{tab13} }

The algebra for 4 dimensions is somewhat similar to the one for 5
dimensions. The global symmetry algebra is $E_{7(7)}$. To extract the
space-time signature from the denominator sub-algebra, we are
dealing with real forms of $A_3$, to be embedded in a real form of
$A_7$. The relevant real forms of $A_3$ are $su(4) \cong so(6)$,
$su^*(4) = so(1,5)$, $su(2,2) = so(4,2)$, $sl(4,\R) = so(3,3)$. Again,
those who are less comfortable with the matrix algebra's are reminded that
also a direct computation is possible. 

\TABLE{
\begin{tabular}{|c|cccccc|c|}
\hline
$\R_{m,n} $ & \multicolumn{6}{c|}{$T_{p,q}$} & $h$ \\
\cline{2-7}
 & $++$ & $+-$ & $--$ & $+-$ & $++$ & $--$ & \\
\hline
$(0,3)$ & $(1,6)$& $(1,6)$& $(3,4)$& $(5,2)$& $(5,2)$ & $(7,0)$ & $so^*(16)$\\
$(1,2)$ & $(0,7)$&        &        &        &         & & $so(16)$\\
$(1,2)$ &        & $(0,7)$& $(2,5)$& $(4,3)$& $(4,3)$ & $(6,1)$ & $so(8,8)$\\
\hline
\end{tabular}
\caption{Dualities of $IIB$-theories in 3 dimensions} \label{tab14} }

In 3 dimensions we have $E_{8(8)}$ as our global symmetry, while the
local symmetry algebra must be a real form of $so(16,\C)$
The embedding of $so(7)$ in $so(16)$ is straightforward. The
easiest way to see it is via the chain of decompositions
\be
so(16) \rightarrow so(7) \oplus so(7) \oplus so(2) \rightarrow so(7)
\oplus so(2)  
\ee
where in the last step again we have selected the diagonal
sub-algebra from the 2 $so(7)$ algebra's. Although a number of entries
in table (\ref{tab14}) can be understood from this decomposition, constructing
the full table requires more accuracy on various signs than this sketchy
argument gives. 

\TABLE{
\begin{tabular}{|c|ccccccc|c|}
\hline
$\R_{m,n} $ & \multicolumn{7}{c|}{$T_{p,q}$} & $h$ \\
\cline{2-8} & $++$ & $+-$ & $--$ & $+-$ & $++$ & $--$ & $+-$ & \\
\hline
$(0,2)$ & $(1,7)$& $(1,7)$& $(3,5)$& $(5,3)$& $(5,3)$& $(1,7)$ && $H_{9}^{n1}$\\
$(1,1)$ & $(0,8)$&        &        &        &        & && $H_9^c$\\
$(1,1)$ &        & $(0,8)$& $(2,6)$& $(4,4)$& $(4,4)$& $(6,2) $&
$(0,8)$ & $H_9^{n2}$\\
\hline
\end{tabular}
\caption{Dualities of $IIB$-theories in 2 dimensions} \label{tab15} }
 
The denominator groups for compactification of $IIB$-theory to 2
dimensions, that are subgroups of the global symmetry algebra $E_{9(9)}$ appear
in table \ref{tab15}. Note that the two possible non-compact algebra's
$H_9^{n1}$ corresponds to torus signatures where a there is an odd number of
space as well as of time dimensions, while $H_9^{n2}$ takes into
account all signatures where there is an even number of space- and
time directions (except for the usual $IIB$-theory on a Euclidean
8-torus, that is covered by the compact form of $H_9^c$).

\TABLE{
\begin{tabular}{|c|cccccccc|c|}
\hline
$\R_{m,n} $ & \multicolumn{8}{c|}{$T_{p,q}$} & $h$ \\
\cline{2-9} & $++$ & $+-$ & $--$ & $+-$ & $++$ & $--$ & $+-$ & $++$ & \\

\hline
$(1,0)$ &  & $(0,9)$& $(2,7)$& $(4,5)$& $(4,5)$& $(6,3)$& $(1,8)$& $(1,8)$& $H_{10}^{n1}$\\
$(1,0)$ & $(0,9)$&        &        &        &        &  & && $H_{10}^c$\\
\hline
\end{tabular}
\caption{Dualities of $IIB$-theories in 1 dimensions} \label{tab16} }

The table \ref{tab16} contains our results for compactifications of
$IIB$-theories to 1 dimension. These algebra's are sub-algebra's of
$E_{10(10)}$. Again, as in the $M$-theory case this table hardly has any
information (beyond the fact that the real form $H_{10}^{n2}$ does not appear).

For compactification to 0 dimensions, or alternatively, any conjecture
on the resurrection of $IIB$-theory from the $E_{11}$ algebra
\cite{Schnakenburg:2001ya} the relevant denominator algebra is
unique. Computations along the lines of \cite{Keurentjes:2004bv} reveal
that it gives all the $IIB$ theories presented in \cite{Hull:1998vg,
  Hull:1998ym}, and no others.  

\section{Conclusions} \label{conc}

In this paper we have extended and improved some techniques
  from \cite{Keurentjes:2004bv}. These provide a firm mathematical
  framework, in which time-like compactifications of supergravities
  can be studied with relative ease. We have rederived and extended
  results of \cite{Hull:1998br, Cremmer:1998em, Hull:1998vg,
  Hull:1998ym} with the aim of elucidating the full duality web for
  the theories introduced in \cite{Hull:1998vg, Hull:1998ym}.

The tables in this paper collect all maximal supergravity theories,
and the dualities between their ultraviolet completions, the $M$- and
type $II$-string theories. A remarkably simple formula (equation
(\ref{sig})) gives the groups that were previously determined by
educated inspection and explicit reduction of higher dimensional
theories. Moreover, our analysis has revealed the possibility
of more groups: $SO(3) \times SO(1,1)$, $SO(3,2) \times SO(3,2)$,
$Sp(4,\R)$, $SL(8,\R)$ can appear as denominator subgroups of $SL(3,\R) \times
SL(2,\R)$, $SO(5,5)$, $E_{6(6)}$ and $E_{7(7)}$. And last but not
least, it allows to do some computations with infinite dimensional
groups, even though the understanding of these, both from the
mathematical as well as from the supergravity viewpoint, is only rudimentary.

Even though we have restricted to $E_{n(n)}$ groups in the
applications, most of the mathematical discussion was phrased in
general terms or can easily be generalized to arbitrary groups. In 
appendix \ref{app} we have computed the groups implied by the
generalization of our formula (\ref{sig}) for arbitrary split
groups. At least some of these should appear when considering
time-like compactification of the theories in \cite{Cremmer:1999du,
  Keurentjes:2002xc}, that are conjectured to be described by the
symmetry algebra's in \cite{Kleinschmidt:2003mf}.

We expect that the developed formalism and insights from it may be
useful to other problems, involving the algebraic structure of
(super-)gravity. In particular their applicability to
infinite-dimensional algebra's provides new tools for computation. We
hope to report on these in the future. 

\acknowledgments

I would like to thank Chris Hull for correspondence.
This work was supported in part by the ``FWO-Vlaanderen'' through 
project G.0034.02, in part by the Belgian Federal Science Policy Office 
through the Interuniversity Attraction Pole P5/27 and in part by the 
European Commission RTN programme HPRN-CT-2000-00131, in which the 
author is associated to the University of Leuven.

\appendix

\section{Possibilities for cosets} \label{app}

In the body of this paper we have restricted ourselves to cosets
$E_n/H_n$ relevant to compactifications of supergravity theories with maximal
supersymmetry. There are however many more examples of
theories involving sigma models on cosets $G/H$, coupled to gravity
\cite{Breitenlohner:1987dg, Cremmer:1999du, Keurentjes:2002xc}, that
can appear in the dimensional reduction of various theories. One can
extend the analysis for these theories to include reduction on one or
more time-like directions. The methods of the present paper can be
straightforwardly extended to these theories.

If we suppose that the group $G$ is a split, finite dimensional simple
Lie-group, then the generalization of formula (\ref{sig}) still
applies. Let $G$ be the split real form, and ${\cal G}$ be a real form
of the complexification of $G$, generated by an inner involution. Then
the possible denominator subgroups $H$ can be easily characterized, by
specifying the signature $\sigma(H)$. The possibilities for $\sigma(H)$
can be directly derived from the signature $\sigma({\cal G})$ and rank
$r({\cal G})$ of the possible ${\cal G}$'s, by: 

\be
\sigma(H) = \frac{\sigma({\cal G}) +r ({\cal G})}{2}
\ee

Given a split $G$, the computation of the possible cosets is now a
simple exercise, that can be easily carried out with the aid of
\cite{Helgason}, or the results collected in
\cite{Keurentjes:2002xc}. The answers are given in table \ref{tabcoset}.

\TABLE{
\begin{tabular}{|l|l|l|}
\hline
$G$ & ${\cal G}$ & $H$ \\
\hline
\hline
$A_{n(n)} = sl(n+1,\R)$ & $su(n+1-p,p)$ & $so(n+1-p,p)$ \\
\hline
$B_{n(n)} = so(n+1,n)$ & $so(2n+1-2p,2p)$ & $so(n+1-p,p) \oplus so(n-p,p)$ \\
\hline
$C_{n(n)} = sp(n,\R)$ & $sp(n,\R)$ & $gl(n,\R)$ \\
                       & $sp(n-p,p)$ & $u(n-p,p)$ \\
\hline
$D_{n(n)} = so(n,n)$   & $so(2n-2p,2p)$ & $so(n-p,p) \oplus so(n-p,p)$ \\
                       & $so^*(2n)$     & $so(n,\C)$ \\
\hline
$E_{6(6)}$             & $e_{6(-78)}$ & $sp(4)$    \\
                       & $e_{6(-14)}$ & $sp(2,2)$  \\
                       & $e_{6(2)}$   & $sp(4,\R)$ \\
\hline
$E_{7(7)}$ & $e_{7(-133)}$ & $su(8)$  \\
           & $e_{7(-25)} $ & $su^*(8)$ \\
           & $e_{7(-5)}$   & $su(4,4)$ \\
           & $e_{7(7)}$    & $sl(8,\R)$ \\
\hline
$E_{8(8)}$ & $e_{8(-248)}$ & $so(16)$   \\
          & $e_{8(-24)}$  & $so^*(16)$ \\
          & $e_{8(8)}$    & $so(8,8)$ \\
\hline
$F_{4(4)}$ & $f_{4(-52)}$ & $sp(3) \oplus su(2)$      \\ 
           & $f_{4(-20)}$ & $sp(2,1) \oplus su(2)$    \\    
           & $f_{4(4)}$   & $sp(3,\R) \oplus sl(2,\R)$ \\                    
\hline
$G_{2(2)}$             & $g_{2(-14)}$ & $so(4) \cong su(2) \oplus
    su(2)$ \\
             & $g_{2(2)}$ & $so(2,2) \cong sl(2,\R) \oplus
  sl(2,\R)$\\

\hline  
\end{tabular}
\caption{Possible cosets $G/H$ for $G$ split. $H$ is defined as ${\cal
    G} \cap G$.} \label{tabcoset}
}

We have also collected various results from our paper in this table
for easy reference.

We stress again that in table \ref{tabcoset} the only ${\cal G}$ that
can appear are those generated by inner involutions; the reader will
look in vain for entries such as $sl(n+1,\R), su^*(n+1)$ and other
real forms that are generated by outer involutions. For the same
reason we only listed $so(2n-2p,2p)$ and not $so(2n-2p-1,2p+1)$; the
latter real forms are generated by involutions that are outer.

These cosets are relevant for reductions including time-like
directions of the theories in \cite{Breitenlohner:1987dg,
  Cremmer:1999du, Keurentjes:2002xc, Keurentjes:2002rc}. At least some
of these cosets are inevitable contained in the theories listed in
\cite{Kleinschmidt:2003mf}, but to decide which ones requires more
detailed computation, that we will not perform here.

\end{document}